\documentclass{article}

\usepackage{PRIMEarxiv}

\usepackage[utf8]{inputenc} 
\usepackage[T1]{fontenc}    
\usepackage{hyperref}       
\usepackage{url}            
\usepackage{booktabs}       
\usepackage{amsfonts}       
\usepackage{amsmath}
\usepackage{dsfont}
\usepackage{nicefrac}       
\usepackage{microtype}      
\usepackage{lipsum}
\usepackage{fancyhdr}       
\usepackage{graphicx}       
\graphicspath{{media/}}     
\usepackage{caption}
\usepackage{subcaption}
\usepackage{mathtools}

\newtheorem{hypo}{Hypotheses}
\newtheorem{rem}{Remark}

\newcommand{\pH}{p$\mathcal{H}$~}

\title{On the Generalization of Data-Assisted Control in port-Hamiltonian Systems (DAC-p$\mathcal{H}$)
}

\author{
  Mostafa Eslami
  , Maryam Babazadeh \\
  Sharif University of Technology \\
  Department of Electrical Engineering \\
  Tehran, Iran\\
  \texttt{\{mostafa.eslami@sharif.edu, babazadeh@sharif.edu\}} \\
}

\begin{document}
\maketitle

\begin{abstract}
 This paper introduces a hypothetical hybrid control framework for port-Hamiltonian (p$\mathcal{H}$) systems, employing a dynamic decomposition based on Data-Assisted Control (DAC). The system's evolution is split into two parts with fixed topology: Right-Hand Side (RHS)- an intrinsic Hamiltonian flow handling worst-case parametric uncertainties, and Left-Hand Side (LHS)- a dissipative/input flow addressing both structural and parametric uncertainties. A virtual port variable $\Pi$ serves as the interface between these two components. A nonlinear controller manages the intrinsic Hamiltonian flow, determining a desired port control value $\Pi_c$. Concurrently, Reinforcement Learning (RL) is applied to the dissipative/input flow to learn an agent for providing optimal policy in mapping $\Pi_c$ to the actual system input. This hybrid approach effectively manages RHS uncertainties while preserving the system's inherent structure. Key advantages include adjustable performance via LHS controller parameters, enhanced AI explainability and interpretability through the port variable $\Pi$, the ability to guarantee safety and state attainability with hard/soft constraints, reduced complexity in learning hypothesis classes compared to end-to-end solutions, and improved state/parameter estimation using LHS prior knowledge and system Hamiltonian to address partial observability. The paper details the p$\mathcal{H}$ formulation, derives the decomposition, and presents the modular controller architecture. Beyond design, crucial aspects of stability and robustness analysis and synthesis are investigated, paving the way for deeper theoretical investigations. An application example, a pendulum with nonlinear dynamics, is simulated to demonstrate the approach's empirical and phenomenological benefits for future research.
\end{abstract}


\section{Introduction}
This work proposes a hypothetical hybrid control framework that decomposes the port-Hamiltonian (p$\mathcal{H}$) dynamics into intrinsic  and extrinsic components, inspired by flight dynamics Data-Assisted Control (DAC) \cite{eslami2024data,eslami2023sequential}. The intrinsic part is handled using nonlinear robust adaptive control via a virtual control input $\Pi$, while a Reinforcement Learning (RL) generate an agent to learn the policy for optimal mapping of virtual control inputs to real actions. The idea is generalized to a broad class of p$\mathcal{H}$ systems, including robotic, electrical, process, molecular, quantum, and energy systems and networks.

p$\mathcal{H}$ systems are an energy-based modeling paradigm well suited to multi-physics and networked systems \cite{kolsch2021}. By construction, they are passive and admit Passivity-Based Control (PBC) methods (interconnection and damping injection) for stabilization and tracking \cite{kolsch2021,krahc2024}. Recent research has explored hybrid control strategies that combine this model structure with model-free learning (e.g. model-free RL \cite{maryam2022rl}) to enhance performance or cope with uncertainty. For example, an actor-critic approach to tune an energy-shaping (passivity-based) controller, preserving stability while optimizing performance, is proposed \cite{sprangers2015}. They showed that actor-critic RL can find near-optimal energy-shaping controllers (difference between stored and supplied energies) that satisfy the \pH system matching conditions. Also similar scheme on a 2-DOF manipulator, demonstrating pendulum swing-up via learned energy and damping terms \cite{nageshrao2014}. More recently,  a time-continuous adaptive feedback controller for general p$\mathcal{H}$ systems, in which the Hamiltonian itself is used as an initial value function for an online learning procedure is proposed
\cite{kolsch2021}. This controller automatically adapts to a given Lagrangian cost, yields asymptotic stability guarantees, and converges to an (inverse) optimal solution. In general, these model-based learning controllers leverage the p$\mathcal{H}$ structure to constrain the search space and ensure passivity: for instance, energy-shaping parameters are learned within the p$\mathcal{H}$ formalism, reducing the need to solve complex PDEs and yielding interpretable policies. Unlike pure model-free RL, they compare policies among \pH-based architectures, e.g. blending known Hamiltonian models with learned performance tuning.

Traditional PBC methods shape the energy and inject damping to achieve stabilization
\cite{sprangers2015}. However, solving the matching Partial Differential Equations (PDEs) involved in energy shaping can be challenging for complex systems. This has motivated combining PBC with adaptive and learning techniques. Several works address uncertainties in p$\mathcal{H}$ models. For example, adaptive control of uncertain nonlinear port-controlled Hamiltonian systems under actuator saturation is studied \cite{wei2011adaptive}. They developed adaptive stabilization laws by exploiting the dissipative Hamiltonian structure and design an \(H_\infty\)-based adaptive controller to handle parametric uncertainties and disturbances. Similarly, various adaptive and robust PBC methods have been proposed (e.g. robust control by operator right inverses, sliding-mode-based schemes, etc.). These approaches typically rely on known system structure and aim to preserve passivity while rejecting disturbances. 

On the other hand, adaptive control has been combined with p$\mathcal{H}$ in an inverse-optimal framework. For general input-affine p$\mathcal{H}$ systems, online learning adaptation for a control-Lyapunov function (chosen as the Hamiltonian) was developed, yielding an explicit feedback law that asymptotically optimizes a given Lagrangian cost \cite{kolsch2021}. This method is hybrid in that it uses the p$\mathcal{H}$ model and passivity-based design as a scaffold, but adapts unknown parts to improve performance. Similarly, other works have used iterative learning (e.g. Bayesian path-following on fully-actuated p$\mathcal{H}$ systems) or primal-dual gradient schemes that preserve the p$\mathcal{H}$ structure for convex optimal control \cite{kolsch2021}. Compared to the traditional model-free RL, these strategies exploit the energy-based model so that learned adjustments need only cover residual uncertainty, not full dynamics.

A growing trend is embedding Hamiltonian structure in neural networks. Port-Hamiltonian Neural Networks (pHNNs) enforce power-conserving structure and have been used for model learning. For example, pHNNs that incorporate inputs and dissipation to model non-autonomous systems is proposed
\cite{desai2021}. In other work the deterministic and stochastic p$\mathcal{H}$ systems is integarted with neural nets, showing how pH-based layers can capture uncertainty in dynamics \cite{PPdi2024integrating}. Also, neural ODE on a Lie group with p$\mathcal{H}$ structure for robot dynamics is suggested that crucially design a combined control that applies energy-shaping plus damping injection on the learned model, ensuring stability of the learned system \cite{duong2024port}. In this hybrid approach, the network guarantees energy conservation by construction and the controller uses classical p$\mathcal{H}$ methods for tracking, so the data-driven model remains safe to use. Overall, physics-informed networks allow gray-box learning of parts of the dynamics while retaining passivity-based control design.

For systems with switching or discrete mode changes, a Gaussian-process regression approach constrained by a switching p$\mathcal{H}$ model call GP-SPHS was proposed \cite{beckers2023}. Here a latent p$\mathcal{H}$ structure and switching policy are learned from data with uncertainty bounds. The GP prior enforces physically plausible energy functions, and the method preserves the compositional (interconnected) nature of p$\mathcal{H}$ components. A hopping-robot simulation demonstrates that GP-SPHS can identify unknown switching dynamics and modes while maintaining energy-based consistency. Although this work focuses on model learning, the resulting model could be used in a hybrid control loop (e.g. robust passivity-based control on the learned p$\mathcal{H}$).

The p$\mathcal{H}$ hybrid control ideas have begun to appear across diverse fields. In robotics, nearly all rigid-body dynamics have a natural p$\mathcal{H}$ form (via Lagrangian mechanics) \cite{kolsch2021}. Recent robot learning papers have exploited this, e.g. a learned p$\mathcal{H}$ model is applied to a quadrotor and design an energy-shaping controller for it \cite{duong2024port}. More generally, embedding p$\mathcal{H}$ structure in robot learning improves generalization and safety (e.g. by preserving energy conservation). In energy and power systems, p$\mathcal{H}$ models are common for multi-domain networks (electricity, gas, thermal) \cite{duindam2009modeling}. These models lend themselves to distributed control by interconnection. Hybrid schemes could learn uncertain network parameters or tune a desired energy function; though explicit RL examples are rare, foundational p$\mathcal{H}$  models have been developed (e.g. gas pipeline networks have p$\mathcal{H}$ formulations). Flight dynamics (UAVs, aircraft) also admit p$\mathcal{H}$ models \cite{fahmi2021port}. The \pH control has been used for helicopter and fixed-wing stabilization (designing controllers as virtual energy-shaping systems). Hybrid learning methods (e.g. adapting damping or feedforward terms) could enhance tracking performance, but detailed studies are scant. 

In molecular dynamics and open quantum systems, the Hamiltonian framework is fundamental. While molecular simulations are typically simulators rather than controllers, recent machine learning methods (Hamiltonian neural networks) aim to capture molecular forces while preserving energy structure \cite{desai2021}. Control of nanoscale or quantum systems often deals with Lindblad-type open dynamics; these can be cast as p$\mathcal{H}$ by treating dissipative interactions as ports \cite{krahc2024}. In principle, hybrid learning could identify unknown Hamiltonians or adapt pulses for quantum control, but this area is largely open. Some works (e.g. Quantum RL) address optimal control of quantum states, and p$\mathcal{H}$ theory offers a physically consistent way to incorporate dissipation.

Data-driven and learning methods have recently been applied to p$\mathcal{H}$ systems, often to complement physical insight with flexibility. The structure-preserving supervised learning for linear SISO p$\mathcal{H}$ systems is studied \cite{pablo2023}. They show that by appropriate parametrization, one can train a neural network to identify or control a p$\mathcal{H}$ system while exactly preserving its energy structure. Such physics-informed learning reduces required parameters and ensures passive behavior of the learned model. Other works include data-driven p$\mathcal{H}$ system identification (e.g. via p$\mathcal{H}$ Dynamic Mode Decomposition \cite{riccardo2023}) and RL for nonlinear PBC, though many such methods assume partial model knowledge or focus on simpler classes (e.g. linear p$\mathcal{H}$ systems, mechanical systems, etc.).

Overall, the reviewed works show that hybrid p$\mathcal{H}$ control unites the rigor of energy-based design with the flexibility of learning. By comparing methods that share the p$\mathcal{H}$ structure (e.g. energy-shaping RL vs. pH-GPs), researchers ensure that learned components respect passivity and thermodynamics. This synergy has been demonstrated in pendulums, manipulators, hopping robots, and more \cite{sprangers2015,beckers2023}. Key takeaways are: (1) Structured learning (embedding Hamiltonian/dissipation priors) leads to more stable and data-efficient models; (2) Adaptive control within p$\mathcal{H}$ (using RL or optimization) can guarantee stability while improving performance \cite{kolsch2021,duong2024port}; (3) Application domains as varied as robotics and energy networks can benefit from such methods because p$\mathcal{H}$ naturally unifies multi-physics interactions \cite{kolsch2021,krahc2024}.

The hybrid framework DAC-p$\mathcal{H}$ extends the Data-Assisted Control (DAC) paradigm by introducing a novel decomposition of p$\mathcal{H}$ dynamics into two structurally distinct components, each addressed with appropriate control strategies. To the best of our knowledge, explicitly splitting general p$\mathcal{H}$ systems in this manner and designing a modular controller that integrates both robust/adaptive and learning-based components is a novel contribution. A detailed mathematical generalization of DAC for p$\mathcal{H}$ systems, along with the benefits of such dynamic decomposition, is presented in Section~\ref{sec:math}. The decomposed system features two distinct parts: a Right-Hand Side (RHS) that represents an intrinsic Hamiltonian flow designed to handle worst-case parametric uncertainties, and a Left-Hand Side (LHS) which acts as a dissipative/input map, addressing both structural and parametric uncertainties. The assignment of "LHS" and "RHS" to these components isn't literal in terms of their physical placement. Instead, it reflects the flexible nature of this decomposition. For instance, the decomposition can be adjusted to create a more adaptable model on the LHS, leaving the remaining static elements to the RHS as residuals. This flexibility is crucial because, in many scenarios, direct access to the physical port is unavailable, and indirect measurements often involve passive elements that introduce small dissipative terms into the LHS. For now, we'll proceed with this understanding. Our primary focus is on developing the DAC-\pH concept. Moving forward, the LHS will be considered solely the conservative part of the \pH system, while the RHS will be purely the dissipative and energy source mapping part to the port.

DAC-p$\mathcal{H}$ fundamentally differs from previous work in RL, adaptive control, and PBC within p$\mathcal{H}$ systems. In DAC-\pH, the most reliable prior knowledge---typically corresponding to the conservative dynamics---is handled using a model-based approach, while the residual dynamics, including dissipative effects and control inputs, are addressed through data-driven learning. Additionally, symbolic, features or structural knowledge can be incorporated into the RL algorithms, as demonstrated in related literature. This framework offers several key advantages across different domains:

\begin{itemize}
    \item \textbf{Reduced complexity for RL:} The decomposition simplifies the RHS learning task, enabling tractable estimation of sample complexity and hypothesis class learnability. This structured reduction accelerates learning, improves accuracy, and enhances prediction confidence by leveraging known affine or sparse dynamics.
    
    \item \textbf{Modular control and policy reusability:} Control of the state variables in the LHS is decoupled from the performance of the RHS, allowing the learned policy to be reused with different closed-loop behaviors by simply tuning LHS parameters---without retraining the RL policy.
    
    \item \textbf{Improved estimation and observability:} The most uncertain dynamics are isolated in the RHS, while the LHS remains fixed. This makes the LHS well-suited for state and parameter estimation, aiding in the treatment of partial observability. The improved estimates can, in turn, enhance the performance of the learning-based RHS controller.
    
    \item \textbf{Robustness via interconnection analysis:} The feedback interconnection between the LHS and RHS subsystems enables the application of small-gain theorems for robustness analysis and synthesis. This provides a natural bridge between Bellman optimality in the RHS and parameter error decay (e.g., temporal difference errors) and Lyapunov-based control in the LHS.
    
    \item \textbf{Towards closed-loop guarantees:} The interconnected system structure presents promising potential for achieving both performance and stability guarantees in closed-loop operation. This claim is supported by phenomenological insights, empirical observations, and preliminary theoretical analysis.

    \item \textbf{Enhanced explainability and safety:} The hybrid architecture facilitates the integration of both soft and hard constraints through the LHS controller, while safety-aware reward shaping or shielding layers can be implemented in the RHS learning process. The presence of the intermediate signal $\Pi$ increases transparency and interpretability of the RL component by exposing internal decision pathways. Furthermore, safety constraints can be systematically enforced via LHS control design, which can explicitly restrict the reachable state sets, contributing to verifiable safety and constraint satisfaction.

\end{itemize}

The rest of the paper is organized as follows.  Section \ref{sec:math} presents the mathematical formulation of general p$\mathcal{H}$ systems and the proposed decomposition. Section \ref{sec:solution} details the decomposition-based control architecture, including the LHS adaptive controller and the RL for RHS. In this section series of hypthesises o the DAC-p$\mathcal{H}$ framework is introduced.  Section \ref{sec:example} outlines an illustrative application of a single pendulum. 

\section{Mathematical Framework}\label{sec:math}
\subsection{General port-Hamiltonian formulation}
The \pH systems provide a unifying energy-based framework for modeling a wide class of physical systems (mechanical, electrical, thermodynamic, etc.), by encoding energy storage, interconnection, and dissipation explicitly \cite{vanderSchaft2017}. A general finite-dimensional p$\mathcal{H}$ system can be written in input-state-output form as
\begin{equation}
\dot{x} = [J(x,\theta) - R(x,\theta)] \nabla \mathcal{H}(x,\theta) + g(x,\theta) u
\end{equation}
where \(x\in\mathds{R}^n\)  is the state, $u\in\mathds{R}^m$ is input, \(\mathcal{H}(x,\theta)\) is the Hamiltonian (total energy), \(J(x,\theta)=-J(x,\theta)^T\)  is the interconnection (skew-symmetric) matrix, the positive definite \(R(x,\theta)\) is the dissipation matrix, and \(g(x,\theta)\) the input map. The output is typically selected \(y=g(x,\theta)^T\nabla \mathcal{H}(x,\theta)\) for power analysis. The p$\mathcal{H}$ systems are inherently (cyclo-)passive: the energy balance \(\dot{\mathcal{H}}=-\nabla \mathcal{H}^TR\nabla \mathcal{H} + y^Tu\leq u^Ty\) ensures passivity if \(\mathcal{H}\geq 0\) \cite{vanderSchaft2017}. This structural property has enabled energy-based control designs (e.g.\ energy-shaping, damping injection) that guarantee global stability and physical interpretability. In addition to the dynamical equations, systems may have additional algebraic constraints $h(x)=0$.

However, practical systems often involve uncertainties and complex dynamics (e.g.\ unknown parameters \(\theta\), unmodeled friction, input nonlinearities). Designing a single control law that is both robust to these uncertainties and achieves high performance can be challenging. Recently, there has been growing interest in hybrid approaches that combine robust/adaptive control with data-driven or learning methods. For example, RL is employed to augment passivity-based control for p$\mathcal{H}$ systems, learning near-optimal energy-shaping controllers via actor-critic methods \cite{sprangers2015}. Meanwhile, adaptive control methods for uncertain p$\mathcal{H}$ systems have been developed (e.g. adaptive stabilization and \(H_\infty\)  control \cite{wei2011adaptive}). Yet, these approaches typically treat the whole system monolithically.

\subsection{Decomposition into LHS and RHS dynamics}
In this paper, we propose a dynamic decomposition that splits the p$\mathcal{H}$ dynamics into two coupled sub-dynamics. One can then design separate controllers for each part: a model-based robust adaptive controller for one part and a data-driven learning module for the other. Specifically, we decompose the dynamics into a LHS Hamiltonian flow and a RHS dissipative/input flow. A new virtual port variable \(\Pi\) is introduced at the decomposition interface. The LHS subsystem is controlled via a nonlinear robust adaptive law that regulates \(\Pi\), estimating unknown parameters and rejecting internal uncertainties. Meanwhile, a RL policy or supervised neural network implements actual control input $u$  as a function of \(\Pi\) and the state, handling uncertainties in the dissipative part and enforcing structure-preserving constraints. This modular architecture (decomposition block + robust controller + learning block) allows each component to be designed and analyzed independently. The proposed splitting of the dynamics is as follows:
\begin{align}
&\text{LHS:}\quad\dot{x} = J(x,\theta)\nabla \mathcal{H}(x,\theta) + \Pi\nonumber\\
&\text{RHS:}\quad\Pi = -R(x,\theta)\nabla \mathcal{H}(x,\theta) + g(x,\theta) u
\end{align}
The LHS is an energy-preserving component, primarily dictated by its inherent physics. Any uncertainty within the LHS is confined to its parameters, denoted by $\theta$, assuming the system's fundamental topology remains constant. For the purposes of this work, we operate under this assumption: the topology is indeed unchanging. The RHS includes dissipation and control inputs and is mapped from virtual input \(\Pi\) using learned or approximated mappings/policy. Here \(\Pi\) acts as a generalized momentum/flow that the LHS subsystem uses as an external input and that the RHS subsystem generates based on  \(u\). In an ideal decomposition, we would design a controller to specify  \(\Pi\) for the LHS subsystem, and then separately solve for the actual input  \(u\) that produces that \(\Pi\). This decomposition is motivated by the observation that \(J(x,\theta)\nabla \mathcal{H}\) represents the conservative interconnection dynamics (with unknown parameters \(\theta\) but no external energy input), whereas \(-R\nabla \mathcal{H}+gu\) contains the dissipative and controlled parts. By controlling \(\Pi\), we effectively control the total LHS energy dynamics. The LHS subsystem (driven by \(\Pi\)) is itself a p$\mathcal{H}$ system with structure \(J\) and Hamiltonian  \(\mathcal{H}\). In particular, its unforced dynamics  \(\dot{x}=J\nabla \mathcal{H}\) preserves \(\mathcal{H}(x)\) (since \(J\) is skew). The term  \(\Pi\) can inject or dissipate energy as needed via the RHS. Hence, we treat  \(\Pi\) as a virtual control input to the LHS. The controller for the LHS will choose \(\Pi\) to achieve some desired energy or state regulation, while a separate module will realize that  \(\Pi\) through the true physical control \(u\). This idea resembles splitting methods or interconnection of subsystems, but here it is an explicit control design strategy. In summary, the overall control objective is achieved by two nested tasks: (i) designing a robust adaptive controller that commands \(\Pi\) to regulate the LHS dynamics (compensating for unknown \(\theta\) and $x$ via adaptation and estimation), and (ii) designing a learning-based mapping/policy from the commanded  \(\Pi\) (and current state) to the actual actuator input \(u\), coping with uncertainties in  \(R\) and \(g\) and preserving the p$\mathcal{H}$ structure.

\begin{figure}
    \centering
    \includegraphics[width=0.85\linewidth]{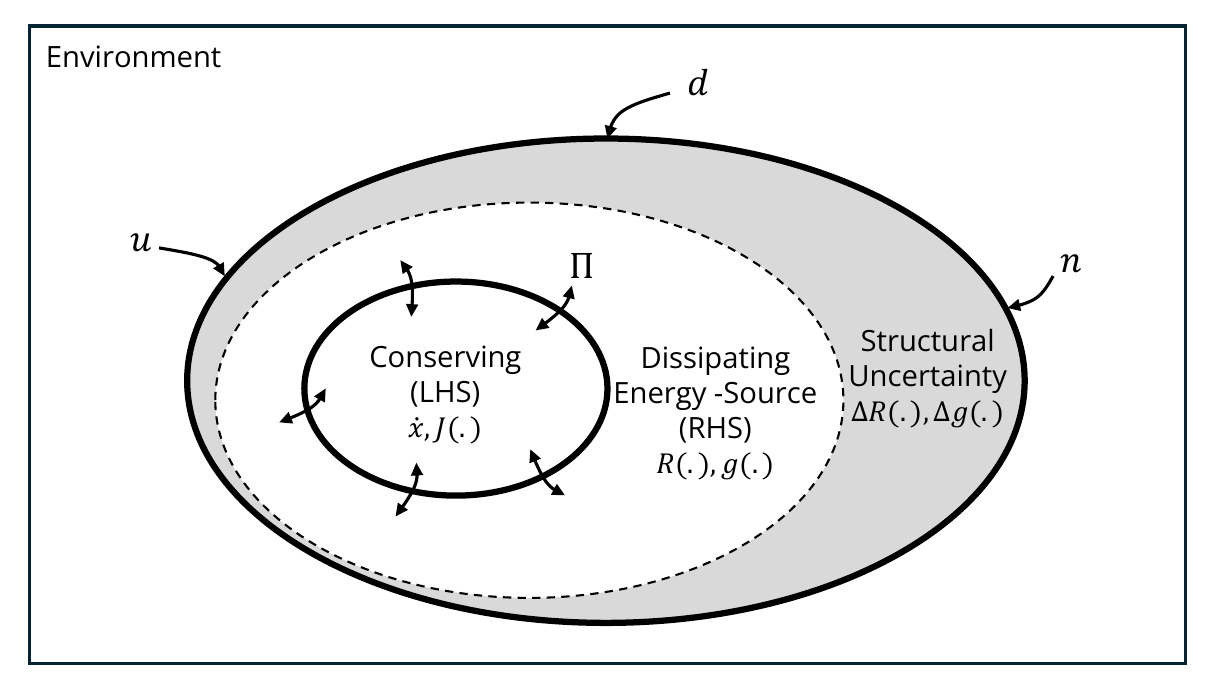}
    \caption{DAC-p$\mathcal{H}$ decomposition, environment, and uncertainty. $d$ is disturbance input and $n$ is noise.}
    \label{fig:pH_model}
\end{figure}

Figure~\ref{fig:pH_model} provides a graphical representation of this decomposition. The innermost dynamics constitute the preserving part, which connects to the dissipative elements and energy sources through the physical port $\Pi$. This preserving part is solely responsible for energy exchange internally, without dissipating or transferring energy to the environment; it houses the system's natural frequencies. We refer to this conserving dynamics as the LHS. While the physical port $\Pi$ exists, it might not be precisely measurable due to inherent dissipations introduced by any measurement process and even infinitesimally small dissipating factors. However, its value can be estimated or calculated using available measurements and superposition principles.

The RHS, on the other hand, comprises the dissipative and energy source elements. Although it has a definitive form, it's highly susceptible to structural uncertainty because it directly interacts with the environment and its models are often truncated or inaccurate. For instance, an exact friction model varies significantly with different cases and environments. Similarly, aerodynamic derivatives in flight dynamics, derived from extensive databases, stem from truncated and imprecise models of environmental interaction. Additionally, unmodeled disturbances and noise, like unforecasted wind gusts, can appear as additional energy-injecting terms in the dynamics, further contributing to the RHS's uncertainty.

Let's examine simple cases to illustrate the defined parts, as depicted in Figure~\ref{fig:pH_examples}. In the electrical system (a), capacitors and inductors form the conserving part. These elements define the system's states (inductor current, capacitor voltage) and primarily exchange reactive energy, leading to phenomena like resonances and defining the system's natural frequency. They are fundamentally non-dissipating. Conversely, the dissipating part includes resistors, which produce heat, and the energy source, which provides input power. The physical ports in this context manifest as capacitor voltage derivatives and inductor currents, enabling the RHS to interact with the environment by dissipating energy as heat or receiving power from the source. While components like capacitors and inductors may have variable values (e.g., $C$ and $L$), the underlying structure of the $J$ matrix remains invariant, unless the circuit's topological layout changes. If non-ideal components, such as series resistors, are present but unmodeled, the physical port location would shift from its ideal representation, and these resistive elements would then be incorporated into the RHS. Environmental factors like temperature also introduce parametric changes, noise, and disturbances, all of which are compatible with this decomposition principle.

In the mass-spring-damper system (b), the mass and spring constitute the conserving part, storing and releasing potential energy in the spring and kinetic energy in the mass. Meanwhile, friction and any external forces make up the RHS. Here, uncertainties primarily stem from the friction model, as its parameters can vary significantly based on environmental conditions, surface properties, and object specifications. Ideally, the port $\Pi$ in this system shares the same dimension as velocity and force. Similarly, the pendulum system (c) also adheres to these boundaries. While the gravitational field influences the storing and releasing of potential energy of the mass, and certain coordinate limits are imposed, its fundamental decomposition aligns. In this scenario, the port dynamics are defined by rotational velocity and torque. We'll further study this system through simulations to test and verify the basic principles proposed in this paper.

\begin{figure}
\centering
\begin{subfigure}{0.3\textwidth}
    \centering
    \includegraphics[width=\linewidth]{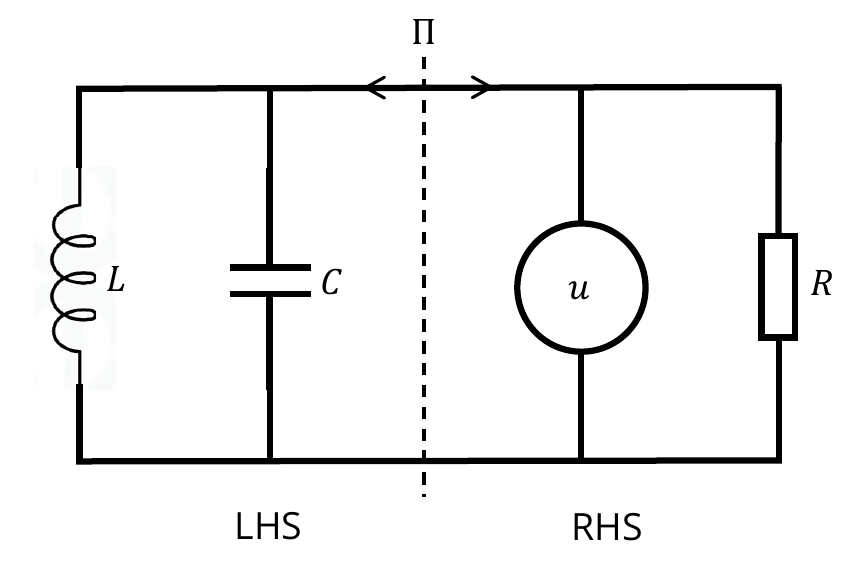}
    \caption{Electrical circuit}
    \label{fig:pH_DAC_citcuit}
\end{subfigure}
\begin{subfigure}{0.3\textwidth}
    \centering
    \includegraphics[width=\linewidth]{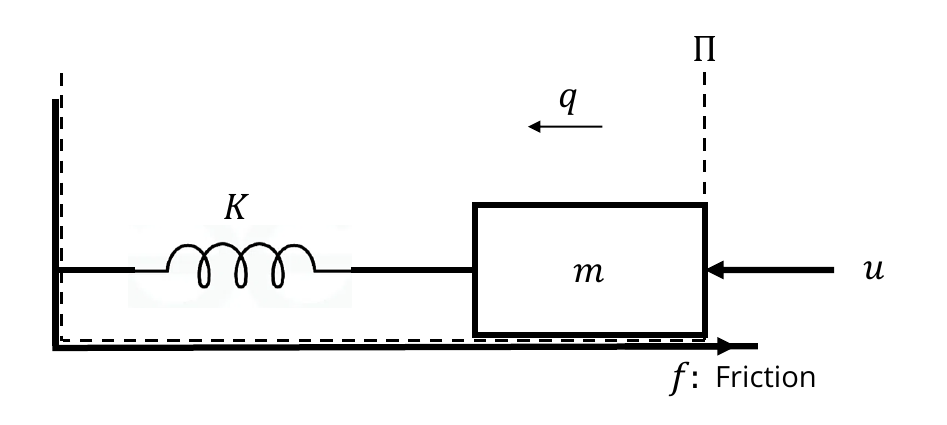}
    \caption{Mass-spring}
    \label{fig:pH_DAC_mass}
\end{subfigure}
\begin{subfigure}{0.3\textwidth}
    \centering
    \includegraphics[width=\linewidth]{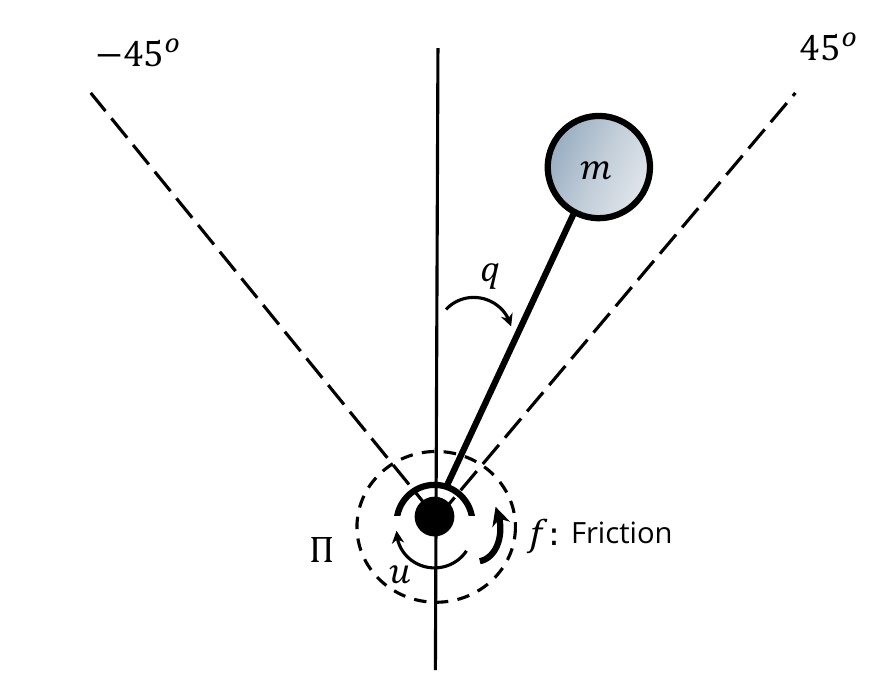}
    \caption{Inverse pendulum}
    \label{fig:pH_DAC_pend}
\end{subfigure}
\caption{Examples of physical system decomposition boundaries and $\Pi$ port location in DAC-p$\mathcal{H}$}
\label{fig:pH_examples}
\end{figure}

\subsection{Modeling language of \pH systems: bond graph}
Following the generalizing approach extends to a wide array of complex, large-scale systems, including vast power grids with numerous electrical components, intricate multi-body and multi-physics systems, swarm robotics, and even molecular dynamics. With certain formal adjustments, it can also be applied to open quantum systems. This methodology integrates seamlessly into cybernetic systems, systems engineering control and optimization tasks, and any system that can be formally modeled and validated using the \pH framework and its related bond graph modeling language \cite{borutzky2011bond,gawthrop2007bond}.

The bond graph methodology stands as a powerful, domain-independent graphical language for modeling dynamic physical systems by explicitly representing power flow and energy exchange between components. This systematic approach naturally leads to the \pH modeling framework, as bond graphs inherently identify the system's energy storage elements, energy dissipation mechanisms, and crucial power-preserving interconnections. This makes bond graphs an ideal tool for deriving p$\mathcal{H}$ models, especially for complex, multi-domain systems. Their ability to unify diverse physical domains under a common power-based language makes them indispensable for understanding and formulating the complex, interacting, and often nonlinear dynamics inherent in multiphysics applications.

Bond graph theory is a domain-independent graphical modeling framework for physical systems, developed to systematically represent energy exchange across multiple physical domains such as mechanical, electrical, hydraulic, thermal, and chemical. It is based on the principle of power continuity, where each connection (bond) transmits power as the product of two conjugate variables: effort (\(e\)) and flow (\(f\)), such that the instantaneous power is \(P = e \cdot f\). The framework uses standardized elements to represent physical phenomena: storage elements (inertial \(I\) and compliant \(C\)), dissipative elements (\(R\)), sources of effort (\(Se\)) or flow (\(Sf\)), and energy transformation elements (transformers \(TF\) and gyrators \(GY\)). These elements are connected via junctions: 0-junctions enforce equal efforts (analogous to nodes in electrical circuits), while 1-junctions enforce equal flows (similar to mechanical series connections).

A key theoretical feature is the assignment of causality, which defines the computational direction of the effort-flow relationships in each bond and determines the differential-algebraic structure of the system equations. Causality helps identify the system's state variables and supports automated derivation of governing equations. Bond graphs reflect the underlying conservation laws (e.g., conservation of energy, mass, charge) and interconnective structure of physical processes. They provide a modular and hierarchical approach to modeling, facilitating system design, simulation, diagnosis, and control. Beyond their use in \pH formulations, bond graphs are extensively applied in mechatronics, systems biology, process engineering, and other domains where multidisciplinary interactions and energy flow must be modeled with physical fidelity \cite{breedveld2011concept}.

Extending bond graph theory to model open quantum systems offers a promising framework for unifying the treatment of classical and quantum dynamics through energy-based, modular representations. In this approach, quantum systems are described in terms of generalized effort and flow variables, where effort may correspond to quantities like Hamiltonian gradients or quantum potentials, and flow to probability currents or expectation-value rates. Classical bond graph elements find natural quantum analogues: storage elements (\(C\) and \(I\)) can correspond to quantum energy storage in modes or observables (e.g., field modes, position/momentum operators), \(R\)-elements can capture quantum dissipation or decoherence via coupling to an environment (as in Lindblad terms), and sources (\(Se, Sf\)) may represent coherent drives or measurement-induced backaction. Gyrators and transformers could model unitary or entangling operations that mediate reversible or cross-domain interactions, while 0- and 1-junctions can encode conservation of probability or commutation structure across interconnected subsystems. Causality, though nontrivial in quantum systems due to measurement and operator noncommutativity, may be interpreted in the context of quantum trajectories or stochastic unravelings. This structure opens the door to graph-based modeling of hybrid quantum-classical systems, consistent with thermodynamic and information-theoretic constraints, while retaining modularity, composability, and physical interpretability inherent in bond graph theory.

Beyond purely physical systems, bond graphs extend their utility to broader applications. Bond graphs can effectively model the tightly integrated physical and computational components. For example, a bond graph can represent a smart grid segment where electrical power flow (physical domain) interacts with communication networks and control algorithms (cyber domain) that manage energy distribution, smart meters, and grid stability. The information flow and control actions can be represented through modulated bonds. While classical cybernetics focuses on feedback and communication, bond graphs offer a foundational physical layer. They can model complex feedback control loops in robotic systems, where the control algorithm (cybernetic element) interacts with the physical actuators and sensors, enabling the analysis of stability and performance across the physical-information boundary. The modularity and hierarchical modeling capabilities of bond graphs are perfectly suited for large-scale systems. This includes, for example, complex manufacturing assembly lines where the power flow through robotic arms, conveyor belts, and pneumatic systems is modeled, along with their interactions. They can also represent large-scale infrastructure systems like water distribution networks, modeling pump stations, pipe segments, and storage tanks. Within a systems engineering context, bond graphs provide a powerful tool for multidisciplinary design optimization and analysis. For instance, in the design of a hybrid electric vehicle, bond graphs can simultaneously model the internal combustion engine, electric motors, battery pack, transmission, and regenerative braking system, allowing engineers to analyze power flows, energy efficiency, and dynamic performance across all domains within a unified framework. This holistic perspective is crucial for identifying design trade-offs and ensuring system-level performance.

Back to the focus of work in this paper, the generalized view of p$\mathcal{H}$ systems is that the  "port" represents an interface for power exchange between two interconnected subsystems or between a subsystem and its environment. It's defined by a pair of conjugate variables: an effort variable and a flow variable. The product of these two variables at any instant represents the instantaneous power flowing through that port. In decomposition of the conserving part (LHS) and the dissipative/source part (RHS), The port $\Pi$ (often denoted as $\Pi=(e,f)$ or similar, where $e$ is effort and $f$ is flow) acts as the energetic boundary between idealized, purely energy-conserving dynamics (LHS) and the parts of the system that dissipate energy or receive external inputs (RHS). Essentially, $\Pi$ is the abstract representation of the "power wires" connecting idealized core system to the real-world elements that account for losses, external drives, and potentially unmodeled effects. 

However, directly measuring the "ideal" physical port $\Pi$ as conceptualized in the p$\mathcal{H}$ decomposition is often impossible in practice, or at least highly challenging, because any real measurement involves some dissipation.  The port $\Pi$ in the p$\mathcal{H}$ formulation of the conserving part implies a perfectly lossless connection. In a physical system, placing a sensor or transducer to measure the variables associated with $\Pi$ will inevitably introduce some impedance, resistance, or damping, thereby altering the very dynamics. Therefore, $\Pi$ must be estimated or calculated using available measurements and using superposition. This is where observers, state estimators, or model-based calculations come into play. You use the available (noisy, dissipative) measurements of system states and inputs, combined with the knowledge of the system's structure, to reconstruct the ideal power flow at the virtual port.

On the other hand, since $\Pi$ represents a physical power exchange, imposing constraints or safety margins on its components (effort and flow) directly translates to limiting physical quantities like maximum force, velocity, voltage, current, or power flow at key interfaces. This makes the safety margins physically interpretable and intuitively meaningful to engineering applications.

\section{Solution Approach Hypotheses}\label{sec:solution}
This work builds upon a series of hypotheses rooted in evidence from previous research on DAC methodologies. These hypotheses will be systematically validated through ongoing theoretical and empirical investigations in subsequent research. The next sections proceed under the phenomenological assumption that these hypotheses hold true.

Before stating the set of hypotheses, some basic definitions from the RHS and LHS control approaches are necessary to be clarified. We separate the hypotheses for three different cases,
\begin{itemize}
    \item Case I: known parameters with full state access, 
    \item Case II:  dynamic (time-varying or uncertain) parameters with full state access,
    \item Case III:  dynamic parameters under partial state observability.
\end{itemize}
In all three cases, we are assuming structural uncertainty on $R$ and $g$ present, as the fundamental assumption of the suggested framework in problem space. The uncertain term $X$ is denoted by $\tilde{X}$, which is equivalent to $X+\Delta X$. The basic terms and preliminary definitions are provided for case I, and they can be generalized for the other two cases. While various learning-based control strategies, policies, or mappings could be applied to the RHS, this work will exclusively focus on RL. In the RL framework of RHS, the state $s$ is defined as set of $\mathcal{S}=\{\Pi,x\}$ and the action $a$ is defined as set of $\mathcal{A}=\{u, K\}$, where $K$ is the parameter that controls the LHS control performance under control law $\Pi_c(K,x,x_d)$. We consider a desired trajectory $x_d$ that the system aims to follow as the ultimate objective. The objective of the LHS in generating $\Pi_c$ is to ensure that $\|x - x_d\| \to 0$. 

In real physical systems, the state \( x \) and its desired values are typically bounded. To ensure safe operation, the state should remain within a specified safe set for all time. If the system is initialized within this set, it is required to stay there, i.e., the set must be forward invariant. We denote this safe set by
\(\mathcal{C}_x = \{x \in \mathbb{R}^n \mid h_x(x, \cdot) \geq 0\},\)
where \( h_x(x, \cdot) \) is a continuously differentiable function that defines the constraint on the state. Also, there are limitations in actuation power and saturation present in control input $u$, i.e. $\|u\|\leq \delta_u$, or $C_u =\{u\in\mathds{R}^m|h_u(u)\geq 0\}$. We assume these two sets are known, as it is a designer parameters and physical limitations that we can achieve through the study of the system. 

These bounds from RHS impose a bound on attainable port parameter $\Pi$ with control input $u$ and trajectory $x$, which we call it port safe set $C_{\Pi_{RL}}$. Just like $x$ case, we define $C_{\Pi_{RL}}=\{\Pi_{RL}\in\mathds{R}^n|h_{\Pi_{RL}}(\Pi_{RL},.)\geq 0\}$. On the other hand, the LHS can request from the controlled-port set $C_{\Pi_c}$ due to defined control law $\Pi_c$,  defined as $C_{\Pi_c}=\{\Pi_c\in\mathds{R}^n|h_{\Pi_c}(\Pi_c,.)\geq 0\}$. For an optimal system design case, the sets $C_{\Pi_{RL}}$ and $C_{\Pi_c}$ should be close to each other, assuming not over- or under-design. But it is not always true. So, there is some discrepancy between these two sets in practice. Since our problem is not system design at the moment, we leave the topic of interest in system design such as sensors, actuator allocations, controllability, observability analysis in synthesis of system and system design optimizations at the moment and focus on control problem, assuming the common safe and attainable set for port as $C_{\Pi}=\{\Pi\in\mathds{R}^n|h_{\Pi_c}(\Pi,.)\geq 0, \;h_{\Pi_{RL}}(\Pi,.)\geq 0\}$.

\begin{figure}
    \centering
    \includegraphics[width=0.75\linewidth]{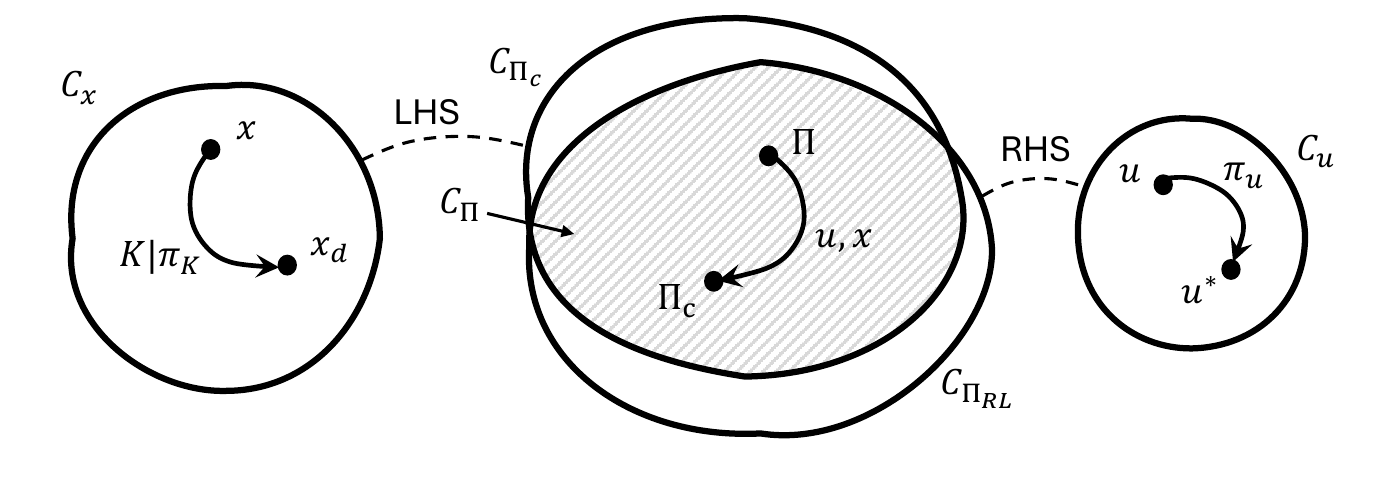}
    \caption{Trajectory of state, input, and port value in DAC-\pH framework}
    \label{fig:pH_DAC_subsets}
\end{figure}

Considering these bound the terminal state of RL in the RHS is defined as $\mathcal{S}^+=\{\Pi_c,x_d\}$, and the RL objective is to through a policy $\pi = \{\pi_u,\pi_K\}=p(a|s)$ maps set of state $\mathcal{S}$ to $\mathcal{S}^+$, such that all state and control input $u$ stay in safe-attainable sets, i.e. $\Pi\in C_{\Pi}$, $x\in C_x$ and $u\in C_u$. Assuming a bounded energy system, which is a reasonable assumption, the sets could be demonstrated by closed subsets, and all these terms can be depicted as Figure~\ref{fig:pH_DAC_subsets}. Additionally, other soft constraints are introduced through subsequent hypotheses and general definitions. These can be handled using reward shaping, which applies penalties for violating them. Hard constraints, however, remain strictly enforced.

Next, hypotheses will be presented. Each hypothesis is supported by relevant theoretical, empirical, literature-based, and conceptual (phenomenological) justifications, all of which are designed to lead directly to rigorous theoretical proofs in subsequent research. The justifications provided are aimed primarily at the most general Case III. However, where current research content is limited, these justifications may be confined to more specific cases.

\begin{rem}
    RL encompasses a broad spectrum of algorithms and theoretical models aimed at enabling agents to make sequential decisions through trial-and-error interaction with an environment. Core distinctions in RL arise across multiple axes: from model-free to model-based methods, from on-policy (e.g., A2C, PPO) to off-policy approaches (e.g., DDPG, TD3, SAC), and with respect to the types of environments addressed—ranging from Markov Decision Processes (MDPs) to partially observable or non-Markovian settings. RL also spans both discrete and continuous state-action spaces and is applied under various task topologies, including hierarchical, multi-agent, or graph-structured systems. Despite this wide diversity, many of these distinctions are secondary in the context of the current work. What remains fundamentally shared across all RL paradigms is that the agent, through approximations such as function approximators or policy gradients, must learn to capture and respond to the underlying dynamics of the environment, even when treated as a black box. The learned policy, value or action-value function implicitly absorbs the temporal and spectral characteristics of the system through reward-driven feedback, enabling the agent to act in a way that effectively controls the system. Thus, RL, irrespective of its formulation, can be seen as an end-to-end framework for learning feedback control strategies through interaction. While RL offers multiple pathways to solving control problems—including planning via model predictive control, explicit model-learning combined with planning or simulation, and policy optimization through direct interaction with the environment—these distinctions are not the focus of the present work. 
\end{rem}

\begin{hypo}
Decomposing system dynamics into a model-based LHS and a learning-based model-free RHS in the DAC-\pH approach reduces the complexity of the hypothesis function class effectively compared to an end-to-end (E2E) approach. 
\end{hypo}

If the hypothesis holds, the DAC-\pH  framework can substantially improve learning efficiency, enhance interpretability and explainability, and provide clearer insights into Out-of-Distribution (OOD) generalization. Additionally, it may enable the derivation of new bounds on the learnability of dynamical systems and significantly narrow the sim-to-real gap. The following presents a set of phenomenological justifications—though not exhaustive—alongside preliminary theoretical analysis to support this hypothesis from multiple complementary perspectives.

\textbf{Phenomenological Justification - General:}
An end-to-end approach must simultaneously address stability, state convergence, implicit parametric dependencies, and state estimation under partial observability. In contrast, a decomposed approach isolates the learning to the RHS, with less complex hypothesis space and target function, without causality dependency of in-out pairs, which simplifies the learning task, ease of learnability and feasibility analysis.

Leveraging DAC-p$\mathcal{H}$ can reduce the function approximation task in RL to a simpler regression problem, potentially avoiding the deadly triad associated with value-based methods in physical dynamical systems \cite{sutton1998reinforcement}. In particular, the estimation of future returns can benefit from a hybrid approach that combines the expected values predicted by the RHS (learning-based) component and those obtained via simulation from the LHS (model-based) dynamics. This fusion can improve the confidence of return predictions and enhance overall algorithmic performance.

During training, the optimization problem becomes more tractable: stochastic gradient descent (SGD) or semi-SGD methods deal with a simpler, affine dependence on the action variables, due to the sparse structure of $R$ and the availability of the  $\nabla \mathcal{H}$. Moreover, the structured decomposition also facilitates the use of non-gradient-based optimization methods, which are now more practical in this context.

\textbf{Phenomenological Justification - RL Optimal Policy:} Assuming structural uncertainty raised from $\tilde{g}(.)$ and $\tilde{R}(.)$ can be identified from experienced data, the ideal control policy within the RL framework should converge to and be expressible as, $\pi_{u,DAC}^\ast(\hat{x},\hat{\theta},\Pi_c,\tilde{g},\tilde{R},\mathcal{H})$, where the estimates $\hat{x}$ and $\hat{\theta}$ are explicitly obtained through the observer and adaptation dynamics defined in the LHS controller. In this case, the hypothesis space for learning the RHS map is static (without temporal features), well-defined and grounded in physical structure: $R(\cdot)$ is assumed to be positive definite, $g(\cdot)$ invertible, and $\mathcal{H}$ differentiable with a known Jacobian, and given control command $\Pi_c$ from LHS.

In contrast, for an end-to-end learning-based control strategy, the ideal policy would be implicitly approximated as $\pi_{u,E2E}^\ast(\hat{x},\hat{\theta},\tilde{g},\tilde{R},\mathcal{H},J,\dot{\hat{x}})$ where entire $\dot{\hat{x}}$, $\hat{x}$ and $\hat{\theta}$ are not available and must be inferred implicitly through learning or treated as hidden parameter. It is a dynamic map (with temporal features) that combines the learning and stability/robustness of the closed-loop system at the same time. The dynamics also include the most complex nonlinearity and coupling inducing term, $J$, which directly impacts the learned policy. In this formulation, the target function to be learned is more complex, and the learning process lacks direct structural constraints. This increases the difficulty of enforcing well-posedness and ensuring stability, leading to a significantly more challenging learning problem compared to the decomposed setting. 

\textbf{Phenomenological Justification - Value Function Approximation:}
In the context of Dynamic-Programming in model-based RL, approximating the value function, $v_{\pi}(s)$, relies on iterative policy evaluation, as expressed by the Bellman optimality equation:

\[v_{k+1}(s) = \mathds{E}_\pi\left[R_{t+1}+\gamma R_{t+2} + \gamma^2 R_{t+3} + \ldots | S_t=s\right]= \mathds{E}_\pi\left[R_{t+1}+\gamma v_{k}(S_{t+1}) | S_t=s\right]\]

Here,  $\gamma<1$ ensures that $v_k(s)$ converges to $v_{\pi}(s)$ as $k\rightarrow\infty$, providing the estimate of future rewards from the current state \cite{sutton1998reinforcement}. Expanding the expected value, we see its dependence on system dynamics:

\begin{align}
    v_{k+1}(s) = \sum_a{\pi(a|s)} \underbrace{\sum_{s',a}{p(s',r|s,a)}}_{System\; Dynamics}\left[r+\gamma v_k(s')\right]
\end{align}

This highlights that accurate policy approximation is intrinsically linked to understanding the system's dynamics. In the DAC-\pH framework, compared to end-to-end RL approaches, the closed-loop dynamics of the LHS are significantly simplified, as demonstrated in Figure~\ref{fig:pH_DAC_subsets}. By design, the LHS controller aims to cancel out most of the complex nonlinear conserving dynamics. Also, the uncertainty that comes from the RHS is common between the two approaches. For instance, the discretized LHS control in Case I will results in, $x_{k+1}=F_k(x_k, x_{d,k},K)$, and $\Pi_{k+1} = G_k(F_k(),u_k)$, where K represents the feedback control law. This architectural choice leads to a closed-loop system with much less complex dynamics such as $F_k = -Kx_{k}+\dot{x}_{d,k}$. In many practical scenarios, especially depending on the time scale of operation, the system's states might even be considered equivalent to their desired values (e.g., $x\approx x_d$ and $\Pi\approx\Pi_c|_{x_d}$, assuming near-perfect tracking. This simplification of the closed-loop dynamics is a key advantage of DAC-\pH, making the value function approximation more tractable and efficient. The precise control laws for various state and parameter availability scenarios are detailed in the second hypothesis. It is straightforward that the same phenomenological justification is valid for value iterations, policy improvement iterations, and hence the full chain of value function evaluations and policy improvement in an iterative setup.

\begin{rem}
Although deriving the justification through dynamic programming and model-based RL can be restrictive, the insights outlined above are, to some extent, generalizable to other value-based RL approaches as well.
\end{rem}

\textbf{Preliminary Theoretical Justification - Complexity of Learning:}
Because the LHS handles known dynamics and state estimation, the RHS only needs to learn the residual or uncertain parts. This constrained structure inherently limits the size and complexity of the function class needed for the RHS. The assumptions on $R(.)$, $g(.)$, and $\mathcal{H}$ further restrict this space. In end-to-end learning, the policy must implicitly learn the system dynamics, state estimation, and control actions simultaneously. This results in a much larger and less structured hypothesis space, potentially encompassing a wider range of complex functions. The PAC (Probably Approximately Correct) framework provides a formal way to define and analyze the learnability of a function class \cite{shalev2014understanding}. A function class is PAC-learnable if there exists an algorithm that, with a reasonable number of samples and computational effort, can learn a hypothesis that is approximately correct with high probability. A function class $\mathds{H}$ is considered PAC-learnable if there exists a learning algorithm that, for any target function $f\in\mathds{H}$, any desired accuracy $\epsilon>0$, and any confidence $1-\delta$ (where $0<\delta<1$), can find a hypothesis $h\in\mathds{H}$ such that the probability of the error of $h$ being greater than $\epsilon$ is less than $\delta$, given a sufficiently large number of training samples $m$ (in continues settings number of samples is equivalent to time of gathering them $T=m\times\Delta t$). A key result in PAC learning relates the number of training samples required to achieve $(\epsilon,\delta)$-accuracy/confidence to the complexity of the hypothesis space. One measure of this complexity is the Vapnik-Chervonenkis (VC) dimension denoted as $VC(\mathds{H})$. The VC measure is valid for labeled data and classification problems, while pseudo-dimensions with real-valued functions and input-output continuous case, denoted as $Pdim(\mathds{H})$ \cite{anthony2009neural}, should be followed here. For a finite $Pdim$ dimension, the sample complexity bound takes the following inequality:
\begin{align}\label{eq:pdim}
  m\geq \mathcal{O}\left(\dfrac{Pdim(\mathds{H})+\log{\dfrac{1}{\delta}}}{\epsilon^2}\right)  
\end{align}

Generally, the exact calculation of the pseudo-dimension $Pdim$ of complex function classes—such as neural networks (NNs) and especially deep neural networks (DNNs)—is not known. However, their qualitative behavior and comparative capacity are relatively well understood through theoretical bounds. For instance, it is shown that the capacity of the network scales polynomially with both the number of parameters and the depth \cite{anthony2009neural,bartlett2017spectrally,anthony2009neural}.

This result implies that while precise $Pdim$ values are typically intractable to compute for specific architectures, the asymptotic behavior and scaling laws are well-characterized. This understanding supports meaningful qualitative comparisons between architectures, guiding insights into generalization and overfitting behaviors.

In the case of DNNs, the depth adds a critical factor to expressivity, often leading to higher capacity and finer function approximation abilities. However, this also increases the complexity of the function class, making exact capacity measures such as $Pdim$ even more difficult to determine. Nevertheless, these networks' performance can still be assessed using upper bounds on complexity measures and empirical generalization behavior.

On the other hand, when prior knowledge of the RHS structure is available, the RL designer can assign an appropriate level of model complexity to the policy and (action-)value networks such that the stability and performance conditions are empirically satisfied. In contrast,  treating the \pH system as a complete black-box requires more complexity in the RL agent, as the observed trajectories play a more significant role in capturing the underlying dynamics, comparing to the DAC-\pH case, where trajectories serve as the primary source of information and are fed into the agent's network as observational inputs

Considering the RHS learning, the hypothesis space is denoted by $\mathds{H}_{RHS}$. Due to the structural assumptions made (e.g., $R(.)$ being positive definite, $g(.)$ invertible, $x$-$\theta$ or their estimate available, potentially parameterized by a neural network with a specific architecture, or belonging to a function class with known properties), one can define a hypothesis space $\mathds{H}_{RHS}$ that contains all possible functions that the learning algorithm can output. The ideal RHS target function $f^\ast_{RHS}$ is the true unknown or uncertain part of the system dynamics that the learning algorithm aims to approximate. Let's assume $f^\ast_{RHS}$ lies within or can be well-approximated by functions in $\mathds{H}_{RHS}$. It is noteworthy that, the target function is equivalent to ideal optimal policy in phenomenological justification. 

In end-to-end learning, the policy $\pi_u^\ast$ is directly approximated. The hypothesis space $\mathds{H}_{E2E}$ would encompass all possible control policies that the learning algorithm can generate. This space needs to implicitly capture system dynamics, estimation, and control actions. The ideal end-to-end policy $f_{E2E}^\ast$ is a complex mapping from the raw state (or partial observations) to the control action.

Because the LHS handles the known dynamics and state estimation, the function that the RHS needs to learn is likely simpler and resides in a smaller, more constrained hypothesis space compared to a function that needs to learn the entire control policy from scratch (as in end-to-end). The structural assumptions on $R(.)$, $g(.)$, and $\mathcal{H}$ further restrict the plausible functions in $\mathds{H}_{RHS}$.  A smaller and more structured hypothesis space generally has a lower VC and pseudo-dimension. For example, if $f^{\ast}_{RHS}$ is parameterized by a neural network with a limited number of layers and neurons, its VC dimension will be finite and potentially smaller than that of a network trying to learn the entire control policy. Accordingly,  lower $Pdim(\mathds{H}_{RHS})$ in the sample complexity bound implies that the RHS learning can achieve the desired accuracy and confidence with fewer training samples compared to learning a more complex function from a larger hypothesis space.

Therefore, given the structural assumptions on $R(.)$, $g(.)$, and $\mathcal{H}$, and the fact that the target function for the RHS is likely simpler (representing only the unknown or uncertain parts), the hypothesis space for the RHS is more likely to be PAC-learnable with a reasonably less number of samples comparing to E2E case. The complex and unstructured nature of the hypothesis space for the end-to-end policy, along with the simultaneous need to learn dynamics, estimation, and control, makes it significantly more challenging to guarantee PAC-learnability within practical sample sizes and computational constraints.

Another approach for qualitative comparison of learning in two cases is using the bias-variance tradeoff. It highlights the balance between a model's ability to fit the training data (low bias) and its ability to generalize to unseen data (low variance). More complex models (like those often required in end-to-end learning for complex tasks) tend to have lower bias but higher variance, making them more susceptible to overfitting, especially with limited data. By incorporating the known system structure into the LHS, the prior knowledge is injected into the learning process. This structured decomposition can act as a beneficial bias, guiding the learning towards more plausible solutions and reducing the risk of overfitting in the RHS learning. On the other hand, the lack of explicit structure in end-to-end learning allows the model to fit the training data very closely, potentially capturing noise and leading to poor generalization (high variance).

\begin{rem}\label{rem:PE_learnability}
    The generalization error bounds derived using $Pdim$ provide a probabilistic guarantee on the difference between the true error (on unseen data) and the empirical error (on the training data) for any function within the hypothesis class. A finite $Pdim$ is a necessary condition for PAC-learnability. However, a low $Pdim$ and a seemingly sufficient number of samples do not guarantee good practical learning if the training data itself is not informative. The strong condition of i.i.d. (independent and identically distributed) sample data practically limits the capability of hypothesis class learning. This is where the Persistency of Excitation (PE) condition (or a similar concept of sufficient input/state variation, such as data augmentation, domain randomization, and data diversification in machine learning theory) becomes crucial, at least for independent sample data generation for RHS. PE, imposed on the LHS dynamics and influencing the control trajectory $\Pi_c$ and the resulting state variations, ensures that the data collected for learning in the RHS is "rich enough" to excite the relevant dynamics and allow the learning algorithm to distinguish between different potential functions within the hypothesis space. Establishing a direct, tight mathematical bound that explicitly incorporates a PE-like condition on the LHS and directly reduces the generalization error bound derived from $Pdim$ for the RHS is a significant research challenge. A constructive approach in Machine Learning Theory, System Identification Theory, Concentration Inequalities, or Information-Theoretic should be followed. The corresponding conditions within the DAC framework are collected and presented in the fourth hypothesis.
\end{rem}

\begin{hypo}
The LHS control law parametrized with gain matrix $K_s(t)$  exists such that the LHS system demonstrates
asymptotic stability, i.e. $\|x(t) - x_d(t)\| \to 0$. Measurements are assumed to be linear combinations of states, i.e., \( y = Cx \). 
\end{hypo}

The LHS dynamics maintain \pH structure, including skew-symmetry of the interconnection matrix \( J(x) = -J(x)^\top \) and structured energy gradients \( \nabla \mathcal{H}(x, \theta) \). These structural properties are exploited to construct Lyapunov robust-adaptive and observer-based control laws. In the case of partial observability, the structural interconnection encoded in \( J(x) \) is assumed to enable the estimation of unmeasured states through topological coupling with measured states, i.e. it is observable. Additionally, according to the fourth hypothesis, the control input \( u \) is designed to ensure PE of the regressor signals necessary for parameter and state estimation.

\textbf{Preliminary Theoretical Justification:} 
In the \pH framework, the LHS dynamics are modeled as:
\[
\dot{x} = J(x, \theta)\nabla \mathcal{H}(x, \theta) + \Pi,
\]
where \( J(x, \theta) = -J(x, \theta)^\top \) by definition, and \( \Pi \in C_\Pi \) is the control input defined to regulate system trajectories. This structure ensures that the internal energy dynamics are passivity-compatible and admits Lyapunov functions in the form of the Hamiltonian.

Under \textbf{Case (1)}, where all state measurements are available and parameters are known and constant, classical nonlinear feedback laws can be synthesized directly from the \pH structure by defining a control law \( \Pi_c \) that ensures closed-loop passivity or desired energy shaping.  For example following control law can be employed,
\begin{equation}
    \Pi_c = -J(x,\theta)\nabla \mathcal{H}(x,\theta)-K_s(x - x_d) + \dot{x}_d 
\end{equation}
Consider a system whose stability can be analyzed using the Lyapunov candidate function:
\begin{equation}
    V_1 = \frac{1}{2}(x - x_d)^\top P(x - x_d)
\end{equation}
Here, $P$ is a symmetric positive definite matrix ($P \succ 0$). This ensures that $V_1 > 0$ for all $x \neq x_d$, and $V_1 = 0$ when $x = x_d$. This is a fundamental requirement for a Lyapunov function, as it acts as a measure of the energy or distance from the desired state $x_d$. The time derivative of $V_1$ is given by:
\begin{equation}
    \dot{V}_1 = \frac{1}{2}\left((x - x_d)^\top P(\dot{x} - \dot{x}_d) + (\dot{x} - \dot{x}_d)^\top P(x - x_d)\right)
\end{equation}
This expression can be simplified to:
\begin{equation}
    \dot{V}_1 =  (x - x_d)^\top (K_s^\top P + PK_s)(x - x_d)
\end{equation}
For the system to be asymptotically stable at $x = x_d$, we require $\dot{V}_1 < 0$ for all $x \neq x_d$. This means that the term $-(K_s^\top P + PK_s)$ must be positive definite, or equivalently, $K_s^\top P + PK_s$ must be negative definite. The expression $K_s^\top P + PK_s$ is directly related to the Lyapunov equation. If we let $W = -(K_s^\top P + PK_s)$, then the Lyapunov equation can be written as:
\begin{equation}
    K_s^\top P + PK_s = -W
\end{equation}
For asymptotic stability, we need $W$ to be a positive definite matrix ($W \succ 0$). A fundamental result in Lyapunov stability theory states that if $P$ is a symmetric positive definite matrix, then $K_s^\top P + PK_s$ is negative definite (meaning $W$ is positive definite) if and only if all eigenvalues of $K_s$ have negative real parts. Therefore, the condition $\dot{V}_1 < 0$ (i.e., $\dot{V}_1 = - (x - x_d)^\top W (x - x_d) < 0$ for $W \succ 0$) holds true if and only if every eigenvalue $\lambda_i$ of $K_s$ satisfies $\text{Re}(\lambda_i(K_s)) < 0$. This ensures that the system converges to the desired state $x_d$ as time goes to infinity, demonstrating asymptotic stability.

We assume that the states and system parameters are perfectly measured. In the presence of measurement errors, the control law can be augmented with an additional robustness term \( \delta_r \) added to the gain matrix \( K_s(t) \). This ensures that sufficient damping is maintained despite the existence of uncertainties.

Under \textbf{Case (2)}, where parameters are uncertain but states are fully observed, adaptation laws (e.g., gradient-based or least-squares estimators) can be designed to estimate parameters \( \theta \). The control law is updated online using \( \hat{\theta} \), and Lyapunov-based stability proofs are enabled by including parameter estimation error dynamics. The key requirement is that the control input \( u \) excites the system sufficiently to render the parameter regressor PE. In the \pH structure, this condition is related to the energy gradients and their interaction with the control input. For example following control law can be employed,
\begin{equation}
    \Pi_c = - J(x, \hat{\theta})\nabla \mathcal{H}(x, \hat{\theta})+K_s(t)(x - x_d) + \dot{x}_d 
\end{equation}
and,
\begin{equation}
    \dot{\hat{\theta}} = \Gamma \phi(x)^\top P(x - x_d) , \quad \Gamma\succ 0,
\end{equation}
where $\phi(x)$ is regressor of the parameters. Therefore, the Lyapunov candidate and its derivative can be derived as follows,
\begin{equation}
    V_2 = \frac{1}{2}(x - x_d)^\top P(x - x_d) + \frac{1}{2}\tilde{\theta}^\top \Gamma^{-1} \tilde{\theta}
\end{equation}
where \( \tilde{\theta} = \hat{\theta} - \theta \). And we have,
\begin{align}
    \dot{V}_2 =  &\frac{1}{2}(x - x_d)^\top P(J(x, \theta)\nabla \mathcal{H}(x, \theta) - J(x, \hat{\theta})\nabla \mathcal{H}(x, \hat{\theta})+K_s(t)(x - x_d))\\
    &\frac{1}{2}(J(x, \theta)\nabla \mathcal{H}(x, \theta) - J(x, \hat{\theta})\nabla \mathcal{H}(x, \hat{\theta})+K_s(t)(x - x_d))^\top P(x-x_d)\\
 &+ \frac{1}{2}\tilde{\theta}^\top \Gamma^{-1} \dot{\tilde{\theta}} +\frac{1}{2}\dot{\tilde{\theta}}^\top \Gamma^{-1} \tilde{\theta}.
\end{align}
Assumption slowly varying $\theta(t)$, $\dot{\theta}\approx
 0$ and hence $\dot{\hat{\theta}} \approx \dot{\tilde{\theta}}$, and regressor matching assumptions, i.e. $J(x, \theta)\nabla \mathcal{H}(x, \theta)- J(x, \hat{\theta})\nabla \mathcal{H}(x, \hat{\theta}) \approx -\phi(x)\tilde{\theta}$; we have,
\begin{equation}
    \dot{V}_2 =  - (x - x_d)^\top W(x - x_d) < 0 .
\end{equation}
An additional robustness term can be incorporated into the control law to compensate for the approximation errors. This control law leads to $V_2(x,\theta)\rightarrow 0$ and hence $x\rightarrow x_d$.

Under \textbf{Case (3)}, where only partial state information is available (i.e., \( y = Cx \)), a state observer is required. Because \( J(x) \) encodes topological interconnections (e.g., in electrical, mechanical, or network systems), the system is often structurally observable—meaning the measured outputs can be propagated through the known topology to reconstruct unmeasured states. Observer design (e.g., Luenberger or nonlinear adaptive observer) can then be employed, and the observer gain \( L \) is tuned to ensure convergence of \( \hat{x} \rightarrow x \). The adaptation laws for \( \hat{\theta} \) are modified to use the estimated states \( \hat{x} \) instead of the true \( x \). Due to the complexity of the design and its connection to the RHS learning-based control, both components should be designed concurrently.

In all three cases, the control law and estimation dynamics can be synthesized from the known \pH form and benefit from its structured decomposition. Furthermore, because \( \Pi_c \) is generated via a model-based process, its evolution is analytically tractable and its interaction with the learned RHS policies is interpretable. This structure enables the derivation of Lyapunov-based guarantees for stability and robustness across different regimes of observability and uncertainty.

\begin{hypo}
The RL policy in the RHS operates on any states $x\in C_x$, the port variables $\Pi\in C_\Pi$, and the control inputs $u\in C_u$ to guide the system towards its terminal state, ensuring full state controllability $\Pi\rightarrow \Pi_c$ and $x\rightarrow x_d$ for any $\Pi_c\in C_{\Pi}$ and $x_d\in C_x$. While it can temporarily exceed soft limits within predefined feasible and safe sets, it's strictly prevented from violating hard limits. 
\end{hypo}

First of all, due to the uniqueness of the solution to the LHS differential equation, the convergence \( x(t) \to x_d(t) \) is equivalent to \( \Pi(t) \to \Pi_c(t) \). This equivalence holds under the assumption that \( J(\cdot) \) is locally Lipschitz in \( x \), and the Hamiltonian \( \mathcal{H}(\cdot) \) is twice continuously differentiable (\( C^2 \)). Under these regularity conditions, the LHS differential equation admits a unique solution for any piecewise continuous and bounded virtual input \( \Pi_c(t) \).

Building on the idea that Hypothesis 2 allows for any desired state $x_d\in \mathds{R}^n$ to be reached by a control law $\Pi_c$ across all three cases, we now consider the critical challenge of achieving full state controllability under explicit soft and hard constraints on the state ($x\in C_x$), the virtual control ($\Pi_c\in C_\Pi$), and the actual input ($u\in C_u$). To manage these constraints, we propose a two-tiered approach. Soft limits are implemented through incentivized control mechanisms like reward shaping, encouraging desirable behavior. Hard limits, on the other hand, are strictly enforced using techniques such as shielding and Control Barrier Functions (CBFs), which actively prevent the system from violating constraints.

A key challenge lies in the dynamic discrepancies between the desired virtual control $\Pi_c$ and the actual virtual control generated by the RHS policies over $K$ and $u$. While reaching $\Pi_c\in C_\Pi$ should theoretically lead to $x\rightarrow x_d$, the system trajectories must remain within the constrained sets. Additionally, the dynamics of the discrepancy error between the RHS and LHS must converge, ensuring that the measure $\|\Pi-\Pi_c\|$ consistently decreases over time.

The feasible and safe set $C_{\Pi_{RL}}$ is defined as the image of the policy-induced map $\Pi = -R(x,\theta)\nabla H + g(x,\theta) u$. This map is generated under a safety-aware control law $u=\pi_u(s)\in C_u$, ensuring the physical input adheres to its limits, and the state $x$ remains within $C_x$. The LHS controller, parameterized by a gain matrix $K$, is designed to minimize the error $e_x=\|x-x_d\|$ (for Case I), which indirectly minimizes the discrepancy $e_{\Pi}=\|\Pi - \Pi_c\|$. If $\Pi$ is measurable or its estimate is available, the RL policy is guided by a reward signal that penalizes this discrepancy and promotes safe actions. Subsequently, a shielding layer, potentially incorporating CBFs, directly forces the states to remain within the constrained sets. The gain matrix $K$ is co-adapted as $K_s(t)$, with its dynamics meticulously designed to maintain the negative definiteness of the LHS Lyapunov function. This adaptation accounts for the evolving RHS policy, potentially relying on assumptions like time-scale separation for theoretical guarantees. 

\begin{figure}
    \centering
    \includegraphics[width=0.75\linewidth]{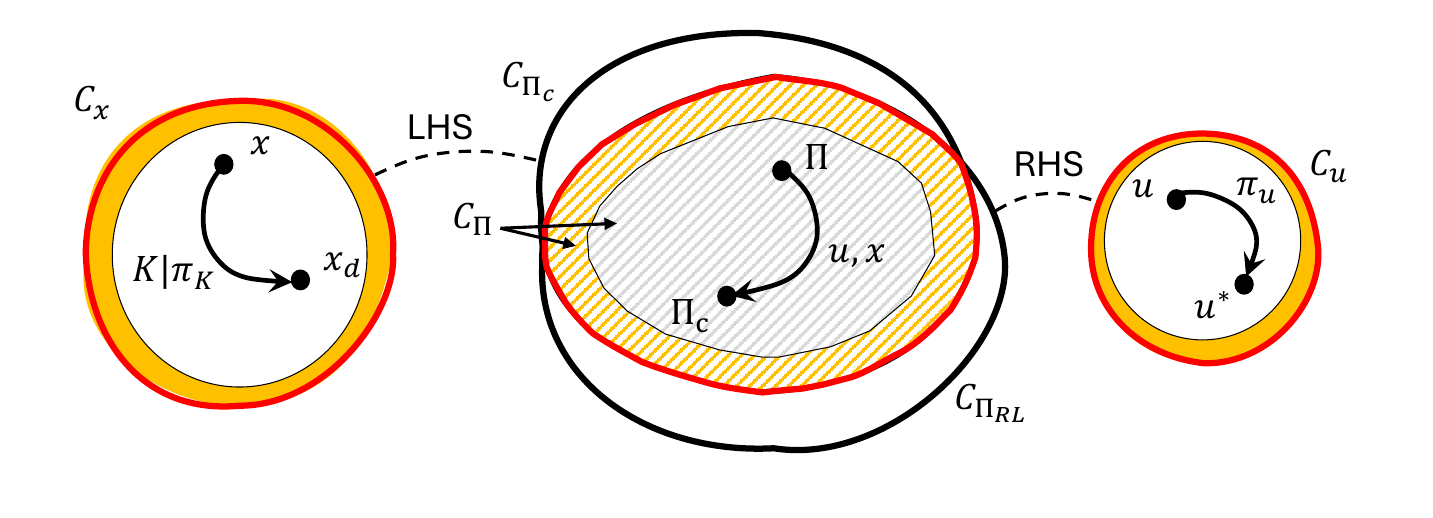}
    \caption{Illustration of soft and hard constraints in DAC-\pH. Red lines denote hard boundaries enforced via CBFs, while the orange region indicates soft safety margins penalized through reward shaping.}
    \label{fig:pH_DAC_subsets_sh}
\end{figure}

\textbf{Phenomenological Justification:} The RL algorithm employs a reward function that balances performance and safety by incorporating soft constraints through reward shaping, as illustrated in Figure~\ref{fig:pH_DAC_subsets_sh}. In the figure, red boundaries represent hard constraints that must never be violated, while the orange region corresponds to soft constraints, which the system states and control inputs are allowed to temporarily enter, though doing so is discouraged.

To guarantee safety, hard constraints are enforced on both system states and control inputs using CBFs. Assume that the barrier functions \( h_i(\cdot) \), for \( i \in \{x, \Pi\} \), are continuously differentiable. The closed-loop discrete-time system evolves according to:
\begin{align}
    x_{k+1} &= F_k(x_k, x_{d,k}, K_k) \\
    \Pi_{k+1} &= G_k(F_k(\cdot), u_k)
\end{align}
Here \( F_k \) represents the closed-loop state dynamics, derived analytically from one of the control laws proposed under Hypothesis 2, \( G_k \) maps the evolved state and current input to the next control output, and the evolution of \( \Pi_{k+1} \) is learned through the RL algorithm and assumed to lie in a neural network (NN) function class. In this system, the LHS controller gain at time step \( k \), denoted \( K_k \), is technically time-varying. However, from standard control practice, such gains cannot be changed aggressively or continuously without risking instability or degraded performance. Therefore, \( K_k \) can be treated as a decision variable in a finite MDP, selected by a RL policy, $\pi_K$, that penalizes rapid variation. Since the introduction of this policy must first be justified through analytical analysis, and appropriate constraints must be carefully defined, we defer detailed treatment of this policy to later works.

Assuming this discrete representation, we employ the discrete-time CBF condition to ensure constraint satisfaction:
\[
h_i(s_{k+1}) - h_i(s_k) \geq -\alpha(h_i(s_k)), \quad \text{subject to} \quad u_k \in {C}_u
\]
where \( s_k \in \{x_k, \Pi_k\} \), \( \alpha(\cdot) \) is a class-\( \mathcal{K} \) function, e.g., \( \alpha(h) = \gamma h \) with \( \gamma \in (0,1) \), and \( {C}_u \) denotes the admissible control input set. To enforce these hard constraints during actuation, a Quadratic Program (QP) is solved at each time step:
\begin{align}
    u_k = \arg\min_{u} \quad & \| u - u_{\text{RL},k} \|^2 \\
    \text{subject to} \quad & h_i(s_{k+1}) - h_i(s_k) \geq -\alpha(h_i(s_k)), \quad i \in \{x, \Pi\} \\
    & u \in {C}_u
\end{align}
Here, \( u_{\text{RL},k} \) is the nominal control action proposed by the RL policy, and the QP ensures that the applied control respects both safety and input constraints.

To discourage the system from entering the soft constraint region (orange area), the reward function is shaped to penalize proximity to constraint boundaries. A representative penalty term can be defined such as:
\[
R_i^C = -\beta \max(0, \delta_i - h_i(s)), \quad \beta > 0
\]
where \( \delta_i \) defines a margin (buffer zone) inside the soft constraint boundary, \( h_i(s) \) evaluates the proximity to the constraint, and \( \beta \) is a positive scalar that adjusts the penalty severity. This formulation encourages the RL policy to remain within the interior of the safe region and avoid the soft constraint zone, effectively providing a soft shield to complement the hard guarantees enforced by the CBF-QP mechanism.

The LHS gain matrix $K_s(t)$ is adapted online to ensure that $\Pi_c$  remains within or near the learnable and safe region of $C_\Pi$. The adaptation dynamics for $K_s(t)$ are designed such that the derivative of the LHS Lyapunov function remains negative definite, guaranteeing closed-loop stability despite the evolving RHS policy. The decomposition into a physically structured LHS and a learning-based RHS allows for interpretability, as the LHS represents the desired behavior based on a known model, and the RHS learns to compensate for uncertainties while respecting the system's physical constraints.

\textbf{Preliminary Theoretical Justification:} To address the first part of the hypothesis—guaranteeing controllability within specified constraint sets—a theoretical approach is adopted. Specifically, based on the preceding phenomenological justification, it is assumed that all relevant variables, including the system states, their desired trajectories and terminal states, the control input, and the optimal control input trajectory, remain within a predefined constraint set. These constraints are shaped by both soft and hard limits imposed by the RL agent.

This section investigates the theoretical guarantees for stability and robustness of the complete DAC-\pH framework under the following conditions: parametric uncertainty in $\theta$, structural uncertainty in $R(\cdot)$ and $g(\cdot)$, and partial observability of the state $x$. It is evident that the learning-based controller in the RHS must cooperate with the estimation and control dynamics of the LHS to ensure convergence.

On the LHS, a tracking control law parameterized by a time-varying gain $K_s(t)$, along with an adaptation law governed by a gain $\Gamma$, and/or an observer-based estimator using a gain $L$, may be employed. The choice depends on the degree of observability and parameter variation. These gains are critical in shaping the closed-loop dynamics and are designed to ensure Lyapunov stability of both estimation and tracking errors. RL reward shaping can be strategically designed to promote negative definiteness of a chosen Lyapunov candidate, enforce Input-to-State Stability (ISS) of the full system, and ensure robustness of the overall controller with respect to modeling uncertainties.

\begin{figure}
    \centering
    \includegraphics[width=0.65\linewidth]{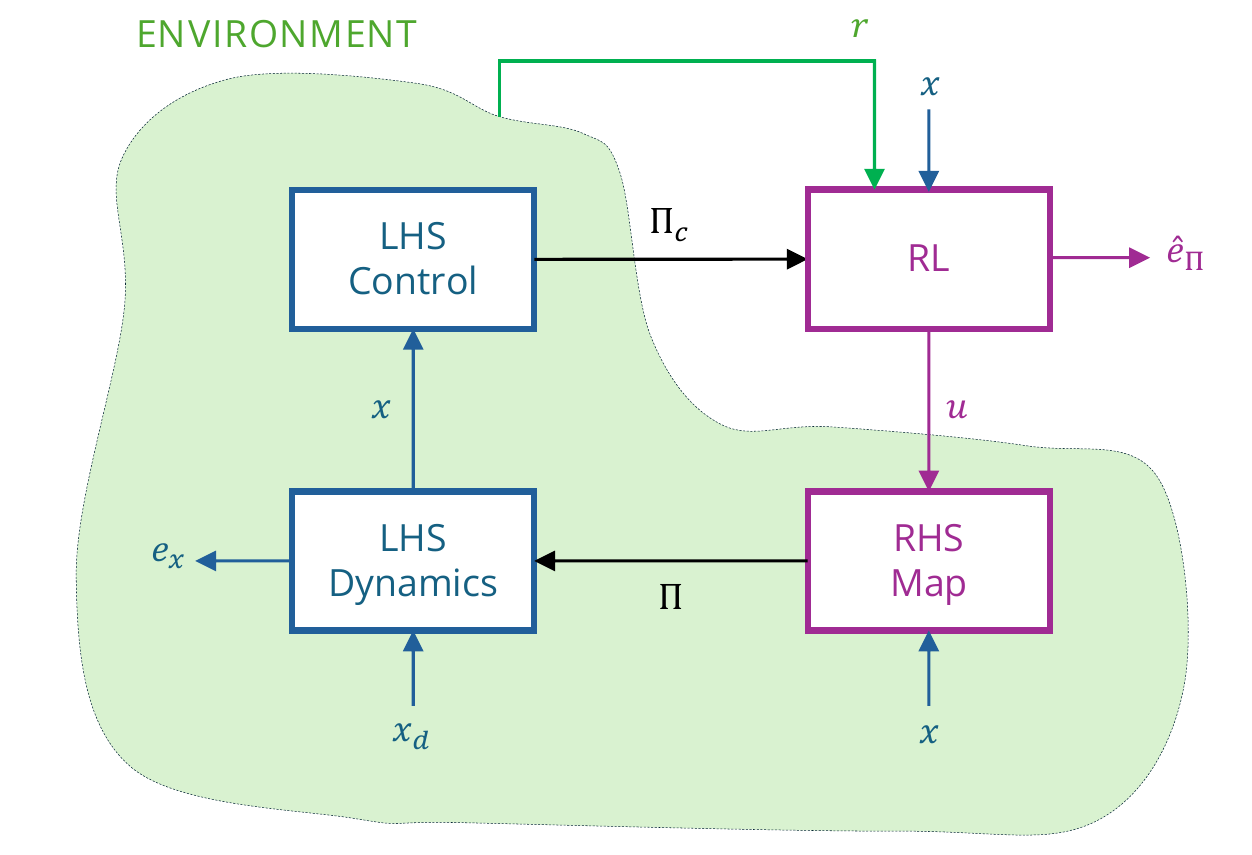}
    \caption{DAC-\pH system architecture - assuming $\Pi=\Pi_c$, the LHS and RHS are virtually closed-loop}
    \label{fig:pH_DAC_closedloop}
\end{figure}

\begin{figure}
    \centering
    \includegraphics[width=0.4\linewidth]{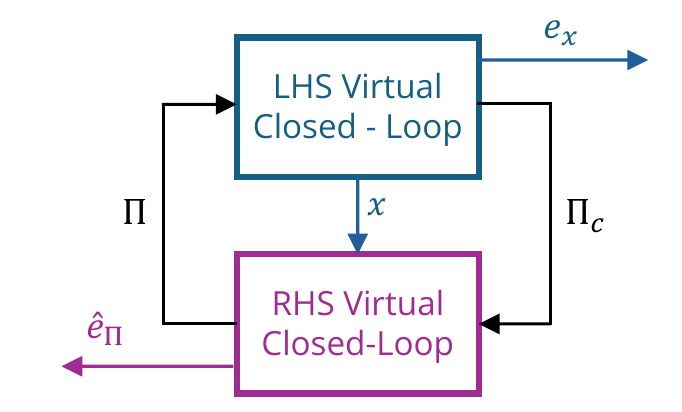}
    \caption{LHS and RHS virtual closed-loop systems in interconnection}
    \label{fig:pH_DAC_md}
\end{figure}

Using the small-gain theorem, stability of the interconnection between the ISS-stable LHS and the RHS learning error dynamics is guaranteed if the loop gain—i.e., from the RHS to the LHS (via $u$) and back from the LHS to the RHS (via measurements or estimates of $x$ and $\theta$)—remains below unity. Figure~\ref{fig:pH_DAC_closedloop} illustrates the full closed-loop DAC-\pH system. Case~I represents the configuration investigated in this preliminary theoretical justification. For clarity and to focus on the core concepts, some connections (such as the gain tuning of $K_s(t)$ via RL) are omitted. A simplified interconnection diagram is also provided in Figure~\ref{fig:pH_DAC_md}.

The LHS closed-loop system has been shown to be ISS, with a Lyapunov function \( V_1 \) such that:
\[
\dot{V}_1 = -W \prec 0,
\]
where \( W \) is positive definite. Consequently, there exist class-\(\mathcal{K}\) functions \( \gamma_1(\cdot) \) such that:
\[
\|x - x_d\| \leq \gamma_1(\|\Pi - \Pi_c\|).
\]

Let the RL policy in the RHS be learned using value iteration or policy iteration. These algorithms are contraction mappings under the Bellman operator \( T^\pi \), implying convergence and error decay. Define the tracking or approximation error as $e_\Pi := \|\Pi - \Pi_c\|$. If the learning-based design ensures that 
\[
\|\Pi - \Pi_c\| \leq \gamma_2(\|\Pi_c\|),
\]
where \( \gamma_2(\cdot) \) is a class-\(\mathcal{K}\) function, then by the small-gain theorem, the interconnection of these ISS subsystems is stable if:
\[
\gamma_2 \circ \gamma_1(s) < s, \quad \forall s > 0.
\]
This ensures that the error \( e_\Pi \) decreases across the loop. If the LHS is approximately linear (due to closed-loop control) and the RHS map is affine in control input \( u \), we can approximate the stability condition in linear terms as:
\[
k_L k_R < 1,
\]
where \( k_L \) is the input-to-error gain of the LHS and \( k_R \) is that of the RHS. Since \( k_L \) is determined by the LHS gain matrix \( K_s(t) \), it becomes a powerful design tool. If \( k_R \) can be estimated from the learned policy and model structure (e.g., affine dynamics), this condition offers a practical stability guarantee. We now define a Lyapunov candidate for the entire DAC-\pH system:
\[
V_{\text{DAC}} = V_1(x) + \frac{1}{2} e_\Pi^T e_\Pi.
\]
The time derivative is:
\[
\dot{V}_{\text{DAC}} = \nabla V_1^T \dot{x} + e_\Pi^T \dot{e}_\Pi.
\]
Since the LHS dynamics receive error due to the imperfect RHS policy, we model:
\[
\dot{x} = J(x,\theta) \nabla \mathcal{H}(x,\theta) + \Pi_c + e_\Pi.
\]
Substituting:
\[
\dot{V}_{\text{DAC}} = \underbrace{\nabla V_1^T \left( J(x,\theta) \nabla \mathcal{H}(x,\theta) + \Pi_c \right)}_{= -e_x^\top W e_x} + e_x^T P e_\Pi + e_\Pi^T \dot{e}_\Pi.
\]

We bound the cross-term using Cauchy-Schwarz and Young’s inequality:
\[
e_x^T P e_\Pi \leq \frac{\|e_x\|_P^2}{2\epsilon} + \frac{\epsilon}{2} \|e_\Pi\|_P^2, \quad \forall \epsilon > 0.
\]

Due to policy improvement in RL is monotonic decrease and Bellman operator is a contraction map, it is assumed that for some class $\mathcal{KL}$ function  $\beta(.,.)$ we have $\|e_\Pi\| \leq \beta (\|e_{\Pi}(0),t\|)$, for all $t \geq 0$. Then according to converse Lyapunov Theorem, there exists class $\mathcal{K}$ function $\alpha(\dot)$ such that $e_\Pi^T \dot{e}_\Pi \leq -\alpha (\|e_{\Pi}\|)$.  Replacing this into inequality, we obtain:

\[
\dot{V}_{\text{DAC}} \leq -\left(\|e_x\|^2_{W} - \frac{\|e_x\|_P^2}{2\epsilon}\right) - \left( \alpha(\|e_\Pi\|) -\frac{\epsilon}{2} \|e_\Pi\|_{P}^2  \right).
\]

If \( \alpha(\cdot) \) dominates the growth of the quadratic term, and \( W \) is sufficiently large, this ensures that \( V_{\text{DAC}} \) is decreasing, implying practical stability of the closed-loop system.

\textbf{Literature Based Justification:} Previous research \cite{eslami2024data} has demonstrated the stability and robustness of a similar dynamic decomposition for flight dynamics, though the RHS was controlled dynamically with a Linear-Quadratic Regulator (LQR). This prior work offered theoretical guarantees for the stability of the hybrid algorithm by leveraging LQR-based Lyapunov functions and LHS Lyapunov-based nonlinear controllers. However, incorporating constrained control inputs—and by extension, constrained virtual forces and moments—introduces significant complexities. This leads to challenging reachability synthesis and analysis problems in flight dynamics, directly impacting aircraft maneuverability, which remains an open research topic. Our current approach, utilizing RL and a static map for the RHS, offers a promising pathway to address these reachability issues and provides a generalized solution for \pH systems. While the foundations for stability and robustness are laid out for each component, achieving comprehensive theoretical guarantees for the entire hybrid system remains a key area of ongoing research.

\begin{hypo}
The PE condition on the LHS can, in theory, be related to the sample complexity and the conditions required to reduce generalization error in reinforcement learning. This relationship directly influences the learnability of the RHS policy and/or the (action-)value function class, potentially reducing the number of required samples or the complexity of the hypothesis class. To ensure this connection is effective, conditions must be established that guarantee compatible and consistent excitation signals.
\end{hypo}

The RHS learning-based controller can incentivize PE conditions through its policy or reward function, thereby promoting accurate state and parameter estimation on the LHS. These PE conditions are essential not only for estimator convergence but also, at least, useful for the learnability of the RHS map. Specifically, the statistical richness of the control input trajectories affects both the excitation of regressors for LHS observers and the sample complexity for generalization in RL. PAC-learnability theory provides guarantees that link the expressiveness of the policy class and the distribution of training data—driven by the LHS trajectories—to the accuracy and confidence of learned RHS policies. 

\textbf{Phenomenological Justification:}
The LHS estimator and observer rely on the availability of regressor signals satisfying the PE condition. These regressors arise from the known topology of the \pH system, known interconnection structure \( J(x) \), and control-affine decomposition allowing \( \Pi \) to drive parameter and state estimate convergence. The standard PE condition in continuous time is given by:
\begin{align}
    \exists\; \alpha_1, \alpha_2, T>0: \quad \alpha_1 I\leq\int_{t}^{t+T}{\phi(\tau)\phi(\tau)^T\;d\tau}\leq \alpha_2 I .
\end{align}
where \( \phi(t) \) denotes the regressor vector constructed from measurable quantities (e.g., \( x, \hat{x}, \theta, \hat{\theta}\)). The PE condition enables exponential convergence of the states estimator (via observer gain \( L \)) and the parameters estimator (via adaptation gain \( \Gamma \)). 

In the RHS, we aim to learn a policy \( u = \pi_u(x, \theta,\Pi_c) \) that generates feasible and safe control actions $u\in C_u$ under following physical map:
\begin{equation}
\Pi_c= -R(x, \theta) \nabla \mathcal{H}(x, \theta) + g(x, \theta) u.
\end{equation}

As explained in the first hypothesis, in the DAC-\pH the policy class is denoted by $\mathds{H}_{RHS}$. It is class of map \( (\Pi_c,x,\theta) \xrightarrow{\pi_u} u\) with the target function $f_{RHS}$ characterized by: structural constraints (e.g., positivity of \( R \), invertibility of \( g \)), target function smoothness and Lipschitz continuity, and prior knowledge of system energy topology. Which in turn affects the actual control input \( \Pi \) that acts on the LHS and thus the trajectories observed. However, PAC-learning theory states inequality \eqref{eq:pdim} is a sufficient condition for learnability if the sample data is independent and identically distributed (i.i.d.).  As highlighted in Remark~\ref{rem:PE_learnability}, the practical success of learning critically depends on the appropriate selection of the hypothesis class \( \mathds{H}_{\text{RHS}} \) and the richness of the collected data. Since the RHS policy is trained on state-action pairs sampled from trajectories generated under the influence of LHS estimators and its control law, ensuring PE in the LHS subsystem improves the distributional diversity of the training data. This, in turn, increases the variance in sampled regressors and covariances, enhances the learnability of the target function, and supports the convergence of the RL policy.

Moreover, this mechanism effectively couples the estimation gains \( L \) and \( \Gamma \) in the LHS with the performance and generalization capacity of the RHS RL agent. Thus, there exists an optimal trade-off between inducing sufficient excitation for identification and learning, while preserving the primary control objective—guiding the system toward a desired terminal state. Balancing these goals is nontrivial and constitutes a foundational open problem in integrating machine learning theory, estimation theory, and learning-based control.

As an initial step toward addressing this challenge, we propose a cross-layer coordination mechanism where the PE condition from the LHS is explicitly embedded in the reward structure of the RL policy on the RHS. This incentivizes control actions that maintain identifiability in the LHS while simultaneously enabling effective learning on the RHS. In other words, the PE condition in LHS, necessitates degree of variances that should be generated through additional term in control input $\delta_u$ by the RL. The term \( \delta_u \) is a bounded and persistently exciting signal of the form:
\[
\delta_u(t) = \epsilon(t) \sum_{i=1}^{M} a_i \sin(\omega_i t + \phi_i),
\]
where \( \omega_i \) are incommensurate frequencies, \( \phi_i \) are random phases, \( a_i \) are amplitude coefficients, and \( \epsilon(t) \) is an optional time-decaying envelope to reduce the perturbation over time. On the other hand, the high $\delta_u$ deviates the states from terminal and hence convergence and trackings. Therefore, there the optimal bound for ensuring PE that can be imposed,, can be encoded in iterative learning polices by RL using reward shaping to explicitly encourage PE by penalizing degenerate or low-rank regressor excitation. To encourage the RL policy to favor control trajectories that excite the system, a reward shaping term based on the excitation energy of the regressor matrix can be introduced:
\[
R^{\text{PE}} = \beta \cdot \text{tr}\left( \Phi_t^\top \Phi_t \right),
\]
where \( \Phi_t \) is a regressor matrix formed by collecting relevant LHS regressors over a time window \( [t - \tau, t] \), and \( \beta > 0 \) is a reward scaling factor. Therefore, the time-lag parameter \( \tau \) can be linked to the size or complexity of the data used in learning. This parameter can be related to the sample complexity \( m \) required for the learnability of the RHS, such that \( m \geq f(\tau / \Delta t) \), where \( \Delta t \) is the sampling interval. This relationship—whether derived theoretically or established empirically—should be appropriately quantified to ensure that the RHS module receives sufficiently rich data for learning, and that the LHS parameters can be reliably estimated.

To ensure the total input remains within feasible and safe bounds, the QP will be revised as follows:
\[
\begin{aligned}
u^* &= \arg\min_{\tilde{u}} \| \tilde{u} - (u_{\text{RL}} + \delta_u) \|^2.
\end{aligned}
\]

This approach ensures both excitation for learning and robustness via safe control enforcement. As the policy converges and parameter estimates stabilize, the perturbation \( \delta_u \) can be reduced (e.g., via \( \epsilon(t) \to 0 \)) to prioritize performance optimality.

\section{Example - Concept Validation}\label{sec:example}
We investigated the general applicability and promising nature of this approach using a nonlinear pendulum system. The dynamics of a pendulum, as illustrated in Figure~\ref{fig:pH_DAC_pend}, can be represented in a \pH framework as follows:
\begin{align}\label{equ:pend_pH}
    \dot{x}-\underbrace{\left(\begin{array}{cc}
        0 & 1 \\
        -1 & 0
    \end{array}\right)}_{J}\underbrace{\left[\begin{array}{c}
         -mgl\sin(x_1)\\
         x_2/ml^2
    \end{array}\right]}_{\nabla\mathcal{H}(x)}=\Pi=-\underbrace{\left(\begin{array}{cc}
        0 & 0 \\
        0 & c
    \end{array}\right)}_{R}\nabla\mathcal{H}+\underbrace{\left(\begin{array}{c}
        0  \\
        1
    \end{array}\right)}_{g}\tau
\end{align}
where $x_1 = q$, $x_2=p$ (momentum), and Hamiltonian is,
\[
\mathcal{H}(x) = \dfrac{x_2^2}{2ml^2}+mgl\cos(x_1).
\]
This model demonstrates that our knowledge of the LHS is reliable only when precise measurements of the mass $m$, gravitational constant $g$, and length $l$ are available. Consequently, the LHS is primarily subject to parametric uncertainty. Notably, all these parameters are directly measurable from the physical components involved in the LHS.

In contrast, the RHS incorporates friction and energy dissipation using a simplified dissipative constant $c > 0$. Accurate measurement of this constant is generally infeasible due to the partially unknown nature of the friction model and its high variability. Moreover, we simplify the actuation by assuming that the control input $\tau$ can be directly applied. However, in realistic settings—such as with an electric motor—the input is voltage, which through a dynamic closed-loop relation, generates the torque $\tau$. Therefore, the RHS inevitably contains both structural and parametric uncertainties. Nonetheless, the dissipative property guarantees that the resistance matrix satisfies $R \succeq 0$.

In this example, the control objective is to drive the system state to $x \rightarrow 0$, subject to a soft position constraint $q \in [-\pi/4, \pi/4]$, as illustrated in the figure, while minimizing energy consumption. The control input $\tau$ is assumed not constrained. Accordingly, a nonlinear controller is first designed for the LHS, followed by the specification of observation and actuation signals for the RL module handling the RHS and overall system behavior.

\subsection{LHS Control}
The LHS nonlinear controller is defined using the following control law $\Pi_c \in \mathbb{R}^2$, which guarantees that $x \rightarrow 0$:
\begin{align}
    \Pi_c = -J\nabla \mathcal{H}(x) + K_s x, \quad 
    K_s = \begin{pmatrix}
        k_1 & k_{12} \\
        k_{21} & k_2
    \end{pmatrix}, \quad \text{with} \quad \forall i, \text{Re}(\lambda_i(K_s)) < 0.
\end{align}
For simplicity, we assume there is no parametric uncertainty in the LHS at this stage, and that $K_s$ is a fixed control gain. Under these assumptions, the control law becomes:
\begin{align}
    \Pi_c = \begin{pmatrix}
        -x_2/ml^2 + k_1 x_1 + k_{12} x_2 \\
        -mgl \sin(x_1) + k_{21} x_1 + k_2 x_2
    \end{pmatrix}.
\end{align}

To ensure that this control law is physically realizable, it must belong to the attainable set $C_{\Pi_{\text{RL}}}$. From the structure of the RHS in Eq.~\eqref{equ:pend_pH}, the first entry of the applied control is always zero, i.e., $\Pi_c(1) = 0$. This condition imposes the constraints $k_1 = 0$ and $k_{12} = 1/ml^2$. Therefore, in order to ensure convergence $x \rightarrow 0$, the gain matrix $K_s$ must be of the form:
\begin{align}
    K_s = \begin{pmatrix}
        0 & 1/ml^2 \\
        k_{21} & k_2
    \end{pmatrix}, \quad \text{with} \quad k_{21}, k_2 < 0.
\end{align}

Assuming the RHS can generate the required control input $\Pi_c$ with bounded error $e_{\Pi}$, while respecting input constraints, the closed-loop dynamics of the LHS become:
\begin{align}
    \dot{x} = K_s x + e_{\Pi}, \quad \text{with} \quad e_{\Pi}(1) = 0.
\end{align}

\begin{rem}
    The gain matrix $K_s$ determines the convergence rate of the state $x \rightarrow 0$. Increasing the magnitude of the gains results in faster convergence and enhances robustness to control errors introduced by the RHS. However, this comes at the cost of higher torque demand. Since torque is saturated, the design of $K_s$ must remain feasible and realizable by the RL module. One possible strategy is to select $K_s$ from a predefined set and treat it as a discrete action space in the RL algorithm, akin to adaptive gain scheduling in traditional control systems.
\end{rem}

\begin{rem}
    In the final algorithm, the gain matrix $K_s$ should ideally be determined by the RL agent via a policy $\pi_K(\cdot)$, due to the complexity and high dimensionality of the dynamics. However, in this simplified example, an analytical design is adopted to validate the theoretical framework and demonstrate the proof-of-concept.
\end{rem}

\subsection{RHS Control}
Here, continues RL design is followed. The action space is motor torque, i.e. $\mathcal{A}=\{\tau\}$ and states space is $\mathcal{S}=\{x,\Pi_c(2)\}$. In order to reach the origin, the following reward is defined:
\begin{align}
    \mathcal{R} = -w_1\|x_1\|^2-w_2\|x_2\|^2-w_3\|\tau\|^2 - w_4f(x_1),
\end{align}
where, $f(x_1)$ is penalty for vaiolating the state constratint such that if $|x_1|\geq \pi/4$ it takes value of $\|x_1-\pi/4\|^2$, and otherwise $0$, and $w_i$ are weightings fornormalizing and scaling the correspoding term. In this example, a simplified reward function is assumed, and the relevant constraints are ignored at the moment. The additional considerations discussed earlier in Hypothesis 4 are not essential for demonstrating the core concepts in this proof-of-concept scenario.

To address the continuous nature of the system, we employ a stochastic RL algorithm known as Soft Actor-Critic (SAC). SAC is an off-policy actor-critic method designed for continuous control tasks, grounded in the maximum entropy RL framework. It enhances the standard expected reward objective by incorporating an entropy term, which promotes exploration and improves robustness. In this simulation, SAC is used in an on-policy configuration to evaluate its behavior under more challenging conditions, specifically when no prior knowledge of the RHS model or its target policy is available.

\begin{figure}
\centering
\begin{subfigure}{0.45\textwidth}
    \centering
    \includegraphics[width=0.7\linewidth]{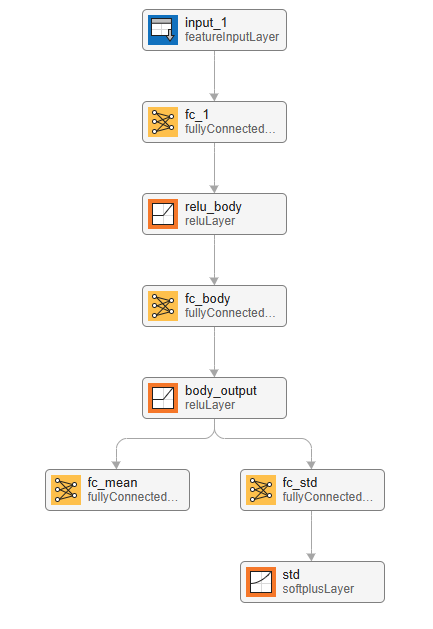}
    \caption{Actor DNN}
    \label{fig:actor}
\end{subfigure}
\begin{subfigure}{0.45\textwidth}
    \centering
    \includegraphics[width=0.7\linewidth]{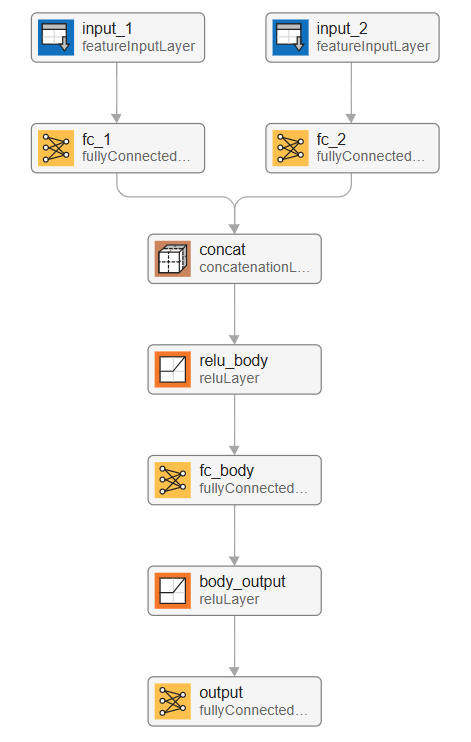}
    \caption{Critic DNN}
    \label{fig:critic1}
\end{subfigure}
\caption{SAC networks' design: Each fully connected input layer (\texttt{fc\_i}) consists of 256 weights and 256 biases per input dimension. Each intermediate fully connected layer (\texttt{fc\_body}) contains \( 256 \times 256 \) weights and 256 biases per input dimension.
}
\label{fig:sac_net}
\end{figure}

For the pendulum example, a deep neural network with the architecture shown in Figure~\ref{fig:sac_net} is employed. Compared to standard configurations commonly used for pendulum control, this network may utilize significantly fewer layers, neurons, and activation functions. However, since the primary goal of the simulation is to demonstrate the proof of concept of the algorithm, network optimization is not addressed in this work.

\subsection{Simulations}\label{sec:sim}
The RHS was trained using random initial conditions with $q(0) \in [-\pi/4, \pi/4]$. The pendulum parameters are listed in Table~\ref{tab:parameter_set}. The LHS control gain matrix $K_s$ is configured to yield eigenvalues $\lambda_1 = -0.5$ and $\lambda_2 = -1$.

\begin{table}
\setlength{\tabcolsep}{5pt}
\renewcommand{\arraystretch}{1.5}
    \centering
        \caption{Parameter set used for simulation of pendulum under DAC control system}
        \vspace{5pt}
    \begin{tabular}{c||c|c|c|c|c|c|c}
        Parameter & $T_s$  & $m$ & $c$ & $l$ & $w_i$ & $k_{21}$ & $k_2$\\\hline\hline
        Value &  $0.05$ [Sec]&  $1$ [kg]& $0.1$ [N.m.s] & $1$ [m] & $1,0.1,0.1,1$ & $-0.5$ & $-1.5$\\
    \end{tabular}

    \label{tab:parameter_set}
\end{table}

\begin{figure}
\centering
\begin{subfigure}{0.48\textwidth}
    \centering
    \includegraphics[width=0.95\linewidth]{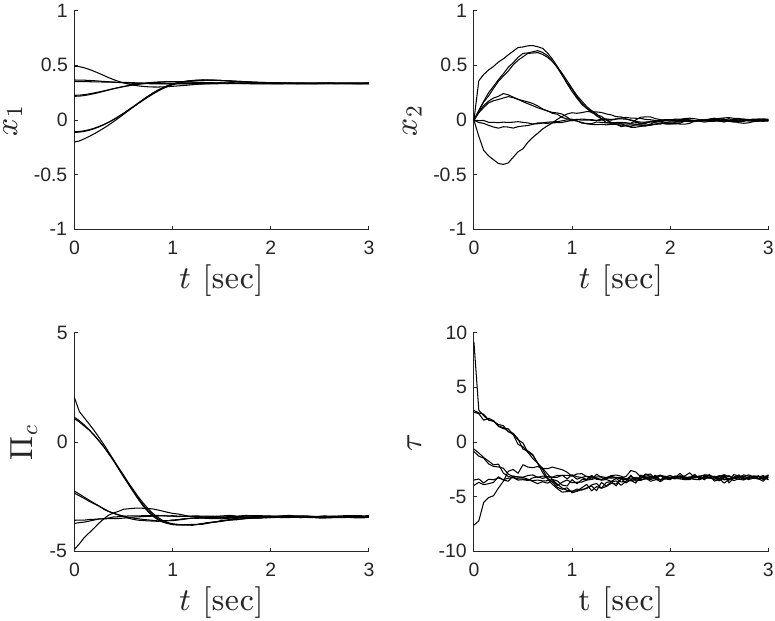}
    \caption{Without integrator}
    \label{fig:10sim_rand_no_int}
\end{subfigure}
\begin{subfigure}{0.48\textwidth}
    \centering
    \includegraphics[width=0.95\linewidth]{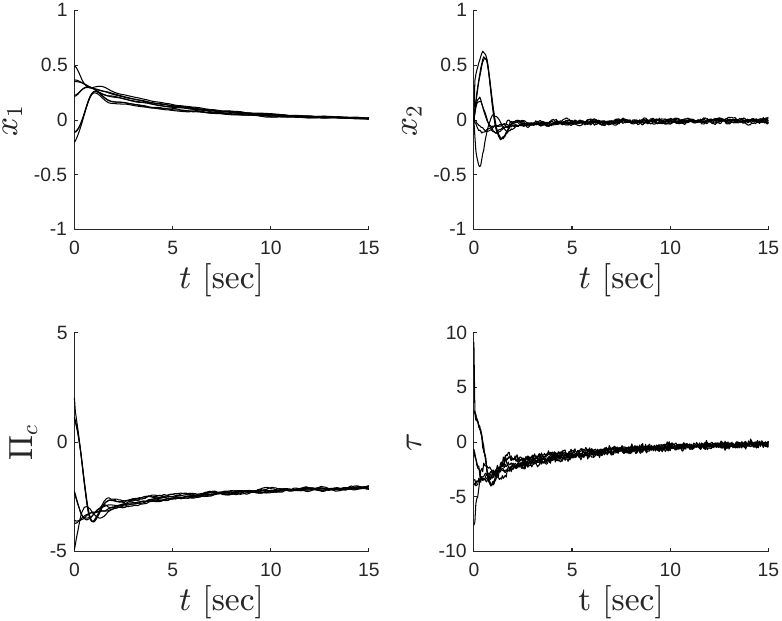}
    \caption{With integrator}
    \label{fig:10sim_rand_with_int}
\end{subfigure}
\caption{Pendulum system under DAC-\pH control: 10 simulations for random initial condition of $x_1\in [-\pi/4,\pi/4]$, $x_{2}=0$}
\label{fig:pH_examples}
\end{figure}

Figure~\ref{fig:10sim_rand_no_int} shows the results of 10 simulations with randomly selected initial conditions, using a state feedback control law. As illustrated, the algorithm consistently demonstrates stable behavior, driving the pendulum to a final position with zero velocity across all trials. Due to residual error in the RHS policy, the LHS closed-loop dynamics experience a disturbance term $e_\Pi(2)$, which introduces a steady-state error in position.

\begin{figure}
    \centering
    \includegraphics[width=0.6\linewidth]{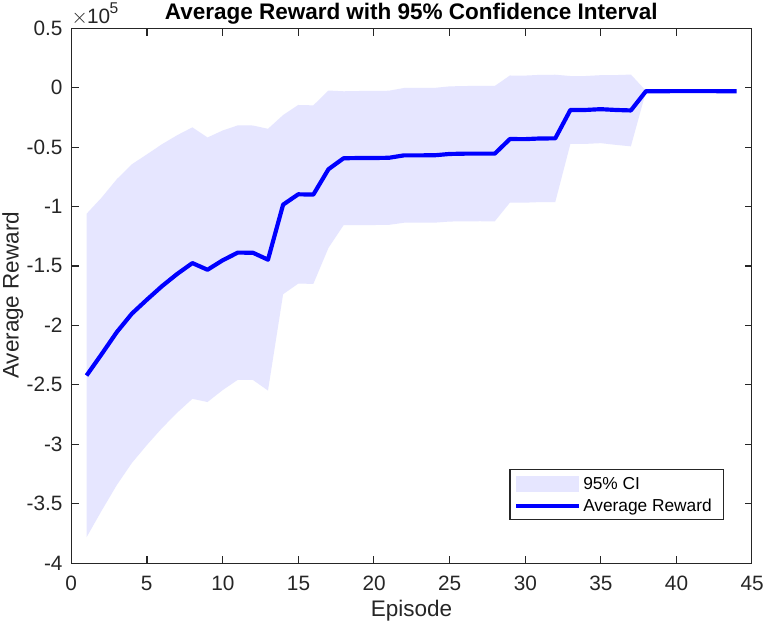}
    \caption{Moving average reward and its corresponding confidence level over training episodes, using a window size of 20}
    \label{fig:moving_reward}
\end{figure}

Nevertheless, the policy reliably tracks the optimal control law $\tau^\ast = \Pi_c + c x_2$. This behavior validates the key concept that the LHS closed-loop dynamics can be tuned independently from the RHS, enabling further refinement through integrator augmentation. Figure~\ref{fig:moving_reward} illustrates the moving average reward for 45 training episodes, calculated with a window size of 20. As shown, it demonstrates a monotonic decrease in penalties.

\begin{figure}
    \centering
    \includegraphics[width=0.45\linewidth]{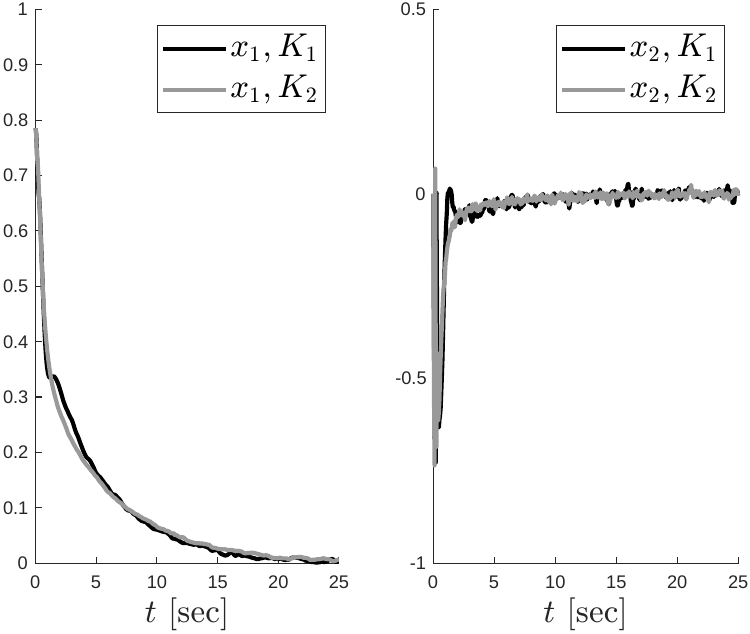}
    \caption{Effect of state feedback gain $k_{21}$ and $k_2$ on overall performance}
    \label{fig:x1x2_k1k2}
\end{figure}

\begin{figure}
\centering
\begin{subfigure}{0.45\textwidth}
    \centering
    \includegraphics[width=0.9\linewidth]{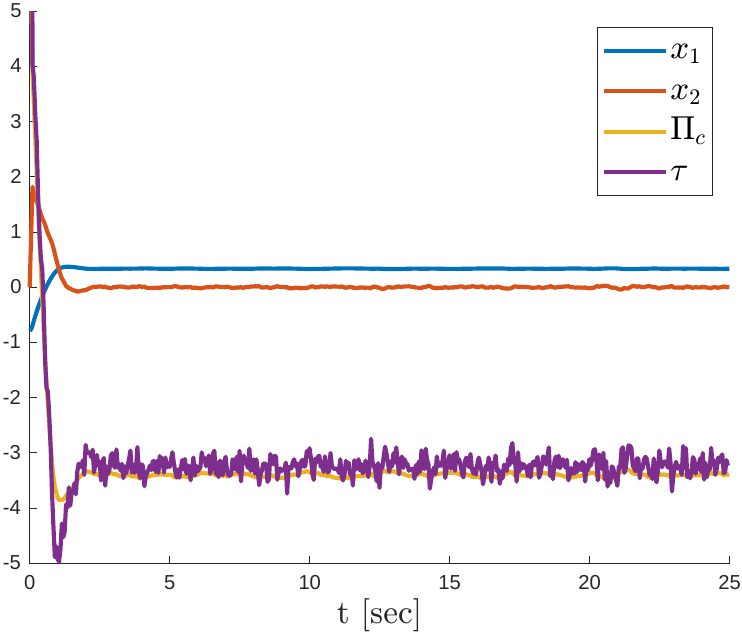}
    \caption{Without integrator}
    \label{fig:sim_no_init}
\end{subfigure}
\begin{subfigure}{0.45\textwidth}
    \centering
    \includegraphics[width=0.9\linewidth]{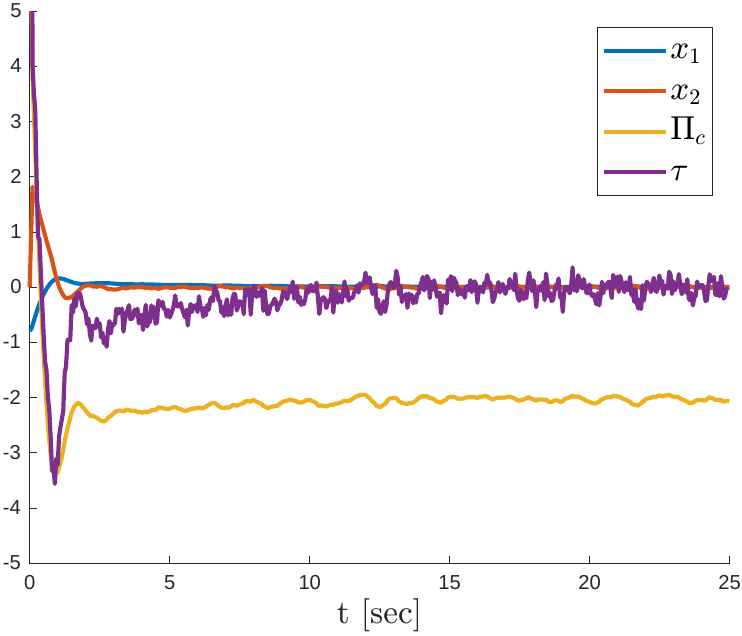}
    \caption{With integrator}
    \label{fig:sim_with_int}
\end{subfigure}
\caption{Effect of LHS integrator}
\label{fig:integrator_effect}
\end{figure}

To eliminate the residual steady-state error, integrators are added to the control law as follows:
\begin{align}
    \dot{z}_1 &= x_1, \nonumber\\
    \dot{z}_2 &= x_2, \nonumber\\
    \Pi_c &= -J\nabla \mathcal{H}(x) + K_s x + K_I z,
\end{align}
where $z = [z_1, z_2]^\top$ represents the integrator states, and $K_i$ is the integrator gain matrix with the negative real part eigenvalues. For integrator enable similations following gain matrix is selected:
\[
K_I = \left(\begin{array}{cc}
   -2  & -1.5 \\
    -1.5 & -2
\end{array}\right)
\]
Figure~\ref{fig:10sim_rand_with_int} presents the simulation results with the integrator-enhanced control law. As evident, the position error now converges to zero. To check the effect of the state feedback gain $K_s$, Figure~\ref{fig:x1x2_k1k2} provides simulation results using a new gain matrix $K_2$ with $k_{21}=-2$ and $k_{2}=-5$. Together, these parameters produce eigenvalues of $\lambda_1=-0.4384$ and $\lambda_2 = -4.5616$. This new gain results in a slightly slower response for state $x_1$ and a much faster response for $x_2$. As both this figure and the previous one imply, the LHS control parameters can be freely chosen, and different control laws can even be implemented without endangering the RHS learned policy for the overall system. Further investigation into performance and robustness analysis is required to check the relationship between the controller gains and stability margins, as well as performance criteria such as overshoot, rise time, and settling times. To investigate the effect of the integrator in more detail, Figure~\ref{fig:integrator_effect} displays the signal changes both with and without its presence. It's evident that without the integrator there's an offset in the control input $\tau$ and $\Pi_c$, consequently affecting the position $x_1$. However, after adding the integrator, $\Pi_c$ increases by around $1 \text{ N.m.}$ to compensate for this bias, which then drives $\tau$ to zero.

At this stage of the research, extensive simulation studies without accompanying rigorous mathematical analysis offer limited value, as the framework is based on a set of interdependent hypotheses that require deeper theoretical investigation. Therefore, the simulations presented here are not intended to validate the entire framework, but rather to provide preliminary evidence supporting key ideas and guide future theoretical development. Despite this scope, several initial findings and supporting observations from simulations have been obtained:

\begin{itemize}
    \item A trajectory-matching mechanism between the RL agent and the LHS controller significantly improves both training speed and accuracy of the obtained policy. In particular, parameters such as the discount factor and reward weights $w_i$ effectively impose a decay rate on the state trajectory, which should be aligned with the eigenvalue design of the LHS gain matrix $K_s$.
    
    \item If the true or estimated value of the intermediate control variable $\Pi$ is available, using the discrepancy error $e_\Pi$ in the reward function (instead of the state error $e_x$) accelerates learning. Due to the uniqueness of the LHS solution, minimizing $e_\Pi$ implicitly minimizes $e_x$.
    
    \item The overall performance and robustness of the closed-loop system depend on the individual performance of its constituent modules. Consequently, appropriate performance metrics should be developed and used for meta-parameter tuning and system-level optimization. Systematic theoretical and empirical methods—such as random sampling and sensitivity analysis—are necessary to establish principled guidelines for hyperparameter tuning.
    
    \item For large-scale systems, leveraging the LHS model to improve reward estimation and predict system trajectories can enhance both learning efficiency and evaluation speed.
    
    \item Among the proposed hypotheses, the first three were empirically tested in this example, and the advantages of the DAC-\pH framework over standard RL algorithms (e.g., those implemented in MATLAB toolboxes) were clearly observed. The fourth hypothesis, concerning the use of $\Pi$ in the reward, requires deeper theoretical analysis and was only partially explored in simulation.

    \item Incorporating an integrator into the LHS control law has proven effective for treating the error signal $e_\Pi$ as a disturbance, thereby enhancing robustness and performance. A dedicated analysis of the general case with integral action—particularly in the context of Hypothesis 2—is necessary to fully understand its implications. Similar considerations may be extended to other control methodologies that interact with the DAC-\pH structure.

\end{itemize}

 \section{Conclusion}\label{sec:conclusion}
This work proposed a general hypothetical hybrid control framework for \pH systems, based on a novel decomposition into internal and external dynamic flows. The DAC-\pH approach adopts a modular design, integrating adaptive control in the intrinsic (LHS) component with RL in the extrinsic (RHS) component. Four fundamental hypotheses were introduced to support this framework, each motivated by phenomenological insight, empirical results, preliminary theoretical considerations, and supporting literature.

The framework was tested through simulations on an inverted pendulum system. Three out of the four hypotheses were successfully validated, and additional promising observations emerged—though these require careful general-case analysis before they can be formalized. The results underscore the potential of DAC-\pH to improve performance, stability, and learning efficiency in complex nonlinear systems.

Future work will focus on providing rigorous theoretical foundations for the proposed hypotheses, transforming them into formally proven theorems. In parallel, deeper integration of prior knowledge into RL formalities—particularly through structural embedding in both the LHS and RHS—will be explored. This includes the development of a novel RL architecture that supports interval-based hybrid model-free/model-based learning. Additionally, extending DAC-\pH to real-world infrastructures such as energy networks (e.g., power grids) using off-policy learning algorithms in closed-loop settings represents a promising and impactful research direction.

\bibliographystyle{unsrt}  
\bibliography{ref,gpt_ref}

\end{document}